\documentclass[aip,reprint]{revtex4-1}

\usepackage{graphicx}
\usepackage{color}
\usepackage{amssymb,amsmath,amsfonts,bm,cancel}
\usepackage[draft]{changes}
\usepackage[normalem]{ulem}
\newcommand{\stkout}[1]{\ifmmode\text{\sout{\ensuremath{#1}}}\else\sout{#1}\fi}

\newcommand{\mb}[1]{\mbox{\bfseries \itshape #1}}
\newcommand{\1}{\Bar{1}}
\newcommand{\2}{\Bar{\Bar{1}}}
\newcommand{\T}{\scriptscriptstyle\rm T }

\newtheorem{defi}{Definition}
\newtheorem{prop}{Proposition}
\newtheorem{coro}{Corollary}

\begin{document}

\title{Observability analysis and observer design of networks of R\"ossler systems}

\author{Irene Sendi\~na-Nadal}
\email{irene.sendina@urjc.es}
\affiliation{Complex Systems Group \& GISC, Universidad Rey Juan Carlos,
28933 M\'ostoles, Madrid, Spain}

\affiliation{Center for Biomedical Technology, Universidad Polit\'ecnica de
Madrid, 28223 Pozuelo de Alarc\'on, Madrid, Spain}

\author{Christophe Letellier}
\email{Christophe.Letellier@coria.fr}
\affiliation{Rouen Normandie Universit\'e --- CORIA, Campus Universitaire du
Madrillet, F-76800 Saint-Etienne du Rouvray, France}


\begin{abstract}
We address the problem of retrieving the full state of a network of 
R\"ossler systems from the knowledge of the actual state of a limited set of
nodes. The selection of the nodes where sensors are placed is carried out 
in a hierarchical way through a procedure based on 
graphical and symbolic observability approaches. By using a map directly
obtained from the governing equations, we design a nonlinear network observer
which is able to unfold the state of the non measured nodes with
minimal error. For sparse networks, the number of sensors scales with half the  network size and node reconstruction errors are lower in networks with heterogeneous degree distributions. The method performs well even in the presence of parameter mismatch and non-coherent dynamics and, therefore, we expect it to be useful for designing robust network control laws.

\end{abstract}

\date{{\it Journal}, to be submitted, \today}

\maketitle

\begin{quotation}
The standard for a real world network is to be nonlinear. This is
observed in biology, medicine, communications, power grid, etc. To investigate,
monitor, control or predict network performance, it is therefore necessary to develop 
appropriate techniques that correctly manage its nonlinear nature. However, most 
of the existing approaches are designed for linear networks and give spurious results
when trying to optimize the placement of sensors in nonlinear ones.
Here we show a procedure that, starting at the node level and
then iterating the dependencies induced by the connectivity, allows to
place the sensors that lead to a global 
observability of the network, that is, to retrieve the whole state of the 
network from a limited set of variables. With this
information, it is then possible to design an observer in the form of a map between the 
gauged variables and all the variables fully describing the network state. That is a prerequisite for an efficient control of 
nonlinear networks.
\end{quotation}

\section{Introduction}

To monitor and control a network of nonlinear dynamical systems, it is
often desirable to be able to determine their state from a limited set of measurements. This is not
a trivial task for many reasons. One of them is that networks from the
real world are often nonlinear by nature, rendering linear approaches inefficient.
There are many examples from technological  (power systems, 
communication systems, traffic) to biological  networks (biochemical 
reactions,\cite{Lev13} brain dynamics,\cite{Sri20} ecological networks\cite{Meng21}).
Among others, the problem results from the complexity of
the considered system, because a network is high-dimensional (due to the node
dynamics and/or the number of interacting units) and nonlinear (each node 
dynamics can be a nonlinear dynamical system and their couplings can be a 
nonlinear function as a sigmoid, for instance). Moreover, there is a limited 
access (number of sensors) to measure the variables required for a full 
knowledge of each state of the network under study.

Observability answers whether a system is observable from a given set of 
measured variables and, to safely reconstruct the state of a system, it is preferable that
they provide a global observability of the state space. The concept of observability was first
introduced for linear 
systems by Kalman.\cite{Kal59} This is a structural observability obtained from
the governing equations and it states whether a system is observable from a 
given set of measured variables or not. This observability was then
extended to nonlinear systems by Hermann and Krener.\cite{Her77} Later on, a 
continuous quantification of observability was introduced for linear and 
nonlinear systems 
through real\cite{Fri75,Agu05,Let98b,Let02} or symbolic\cite{Let09,Bia15,Let18}
coefficients based on Lie-algebraic formulations, allowing to rank which 
variables are the cause for a lack of observability. However, any approach based
on the algebraic computation of the so-called observability matrix to assess 
the observability of a dynamical network, is not applicable due 
to the computational burden of high-dimensional systems, even for small 
nonlinear networks.\cite{Let18} Consequently, a graph-theoretical perspective 
of observability, as developed by Lin,\cite{Lin74} has been presented to tackle networks of  
linear systems.\cite{Liu11,Liu13,Liu15,Agu18,Mon20} The graph is in fact a 
fluence graph, as introduced by Mason,\cite{Mas53} which encodes the 
dependencies (links) between the state variables (nodes). From the structure of 
this graph in terms of the root strongly connected components,\cite{Liu11} it 
is possible to obtain a minimum set of 
variables (sensors) that render the network observable. This approach however 
disregards the nodal dynamics and nonlinearities are not properly considered: 
as a result, it can underestimate the number of sensors as shown by
a series of papers\cite{Mot15,Wha15,Hab18,Let18,Jia19,Hab21} and,
purely graph-theoretic observability approaches may not be
sufficient to design a stable state observer.

While the observability assessment provides information about which output
variables should be measured, it does not tell us how to reconstruct
the network state from those measurements. This is the problem of the
observer design, that is, to unfold the unobserved variables to non-ambiguously retrieve the 
state of the system with the minimal error. While observers for linear systems 
have been successfully uncovered,\cite{Lue64} this is still an open challenge 
for nonlinear systems.\cite{Bes07,Bou21} Several attempts have been proposed 
like using parameter identification,\cite{Wen07,Ang12} global 
modeling,\cite{Agu09} reservoir computing,\cite{Pat17} or through 
optimization-based approaches for jointly observing the states of nonlinear 
networks and optimally selecting the observed variables.\cite{Hab18,Hab21} 

Here, we propose a framework for observing the state of nonlinear dynamical 
networks which uses an observer directly obtained from the governing equations 
of the nodal dynamics through derivative coordinates and a set of sensors 
placed with the help of a procedure combining graphic and symbolic 
approaches.\cite{Let18} Without loss of generality, we exemplify our 
methodology by considering networks of diffusively coupled R\"ossler 
systems.\cite{Ros76c} The knowledge gathered from the observability analysis of 
dyads and triads of R\"ossler nodes\cite{Sen19} is used to propose some rules 
to handle 
larger networks in a systematic way by decomposing the networks in blocks whose 
observability properties are known. Our network observer requires a rather 
short observation horizon --- the duration of time series used to determine
each state of the system --- and it successfully monitors the network state in 
real time as long as the selected sensors are feeding the observer with a 
sufficient time resolution. 

The organization of the manuscript is as follows. In Section \ref{background}, 
we present the formulation of the problem and theoretical background reviewing 
the concepts of observability matrix, observability symbolic coefficients and
the graph approach needed to assess the observability of the nodal
dynamics. In Section \ref{observability} we extend the nodal
observability to the observability of pairs and triads incorporating
the coupling function and present some propositions on the
observability of an arbitrary network. In Section \ref{observer} we
define the observer problem and explicitly construct the observers of
a single R\"ossler unit, and of a pair of them while Section
\ref{networkobserver} is fully devoted to the observer of a network of R\"ossler
systems  and discuss its performance under different network and
dynamical conditions. Finally, we present some conclusions and future work.

\section{Background}\label{background}

\subsection{Problem formulation}

We consider a dynamical network composed of $N$ nodes, each one of them having 
a $d$-dimensional dynamics, whose interactions are given by an adjacency
matrix $A$. We can thus distinguish three levels of description of a
network: (i) the nodal dynamics through the $d\times d$ Jacobian matrix ${\cal
J}_{\rm n} $, (ii) the topology described by the $N \times N$ adjacency 
matrix $A$, and (iii) the whole dynamical network described by the  
$d \cdot N \times d \cdot N$ network Jacobian matrix ${\cal J}_{\rm N}$. 

The node Jacobian matrix ${\cal J}_{\rm n}$,  computed from the set of the $d$
differential equations governing the node dynamics, allows an easy construction
of a fluence graph describing how the $d$ variables of the node dynamics are
interacting. Such fluence graphs were used by Lin for assessing
the controllability of linear systems \cite{Lin74} and, later on, the theory was
extended to address their observability.\cite{Cha92} When nonlinear systems are
considered, it was shown that those edges with a nonlinear component have to be removed from the
fluence graph.\cite{Let18b} When dealing with dynamical networks, it is 
important to distinguish the adjacency matrix $A$ from
the network Jacobian matrix ${\cal J}_{\rm N}$ since, very often the
observability of a network has been wrongly investigated by only taking into
account the adjacency matrix \cite{Bia17,Has13,Mie09} and, thus, disregarding 
the nodal dynamics. 

When the network is actually observable, the observability analysis should identify a candidate set of sensor nodes and state variables which guarantees the determination of the complete 
network state. Then, it remains to express the state of the system in terms of
the original state variables from the measured ones. In the dynamical 
systems theory, it is often sufficient to reconstruct the dynamics by
using derivative or delay coordinates,\cite{Tak81} the so-called 
{\it reconstructed} variables: there is no need to know the governing equations
to do that. Nevertheless, this is not sufficient in some applications
such as the design of a control law where a replica of the real 
system is often required.\cite{Let22a} Therefore, we 
must design a {\it state observer} whose output 
closely follows the evolution of the variables involved in the equations 
governing the system dynamics.\cite{Kre85}

\subsection{Graphical and symbolic observability}
\label{obserap}

Here we briefly present some concepts from observability theory needed to optimally 
select the sensors in nonlinear networks. Let 
\begin{equation}
	\label{system}
  \Sigma \equiv
  \left|
    \begin{array}{l}
      \dot{\mb x} = {\bm f} ({\bm x}) \\[0.1cm]
      {\bm s} = {\bm h} ({\bm x})
    \end{array}
  \right.
\end{equation}
be a dynamical system $\Sigma$ in the state space $\mathbb{R}^d ({\bm x})$ with 
a flow $\phi_t: \mathbb{R} \times {\bm x} \mapsto {\bm x}$ evolving on a smooth 
state space manifold ${\cal M}$ according to the nonlinear vector field 
${\bm f}: \mathbb{R}^d \mapsto \mathbb{R}^d$. The trajectory 
$\phi_t (\mb{x}_0)$ depends on 
the initial conditions $\mb{x}_0$ at time $t_0$. Let be $m$ the number of
variables measured according to the measurement function ${\bm h}: 
\mathbb{R}^d \mapsto \mathbb{R}^m$.

The observability of a system can be defined as follows.\cite{Kai80}
Let us consider the case $m=1$ (a generalization to larger $m$ is 
straightforward), and let ${\bm X} \in \mathbb{R}^d$ be the vector spanning the 
reconstructed space obtained by using the $(d -m)$ successive
Lie derivatives of the measured variables. The dynamical system (\ref{system})
is said to be {\it state observable} at time $t$ if every initial state
${\bm x}_0$ can be uniquely determined from the knowledge of a finite time
series $\{ {\bm X} \}_{t_0}^t$. In practice, it is possible to test whether the 
pair $\lceil {\bm f}, {\bm h} \rceil$ is observable by computing the rank of 
the observability matrix,\cite{Her77} 
\begin{equation}
  {\cal O}_{X} ({\bm x}) =
  \left[
    {\rm d} h({\bm x}), {\rm d} {\cal L}_{{\bm f}} \, {\bm h} ({\bm x}), \dots, 
    {\rm d} {\cal L}^{d-1}_{{\bm f}} {\bm h}({\bm x})
  \right]^{\T} \, .
\end{equation}
where ${\cal L}_{{\bm f}}\, {\bm h}({\bm x})$ is the Lie derivative of 
${\bm h}({\bm x})$ along the vector field ${\bm f}$. The $k$th order Lie derivative 
is given by
\begin{equation} 
{\cal L}_{{\bm f}}^k {\bm h}({\bm x})=\frac{\partial {\cal L}_{{\bm
      f}}^{k-1}{\bm h}({\bm  x})}{\partial {\bm  x}}{\bm f}({\bm x}) \, ,
\end{equation}
${\cal L}_{\bm f}^0\, {\bm h} ({\bm x})={\bm h} ({\bm x})$ being the zeroth 
order Lie derivative of the measured variable. The observability matrix 
${\cal O}_X ({\bm x})$ is the Jacobian matrix of the coordinate transformation
$\Phi: \mathbb{R}^d \mapsto \mathbb{R}^d$ between the state space 
$\mathbb{R}^d (\mb{x})$ and the space $\mathbb{R}^d ({\bm X})$ reconstructed
from the $m$ measured variables.\cite{Let05a} By construction, the 
observability assessed from the observability matrix is a structural 
property.\cite{Agu18}

\begin{defi}
The pair $\lceil {\bm f}, {\bm h} \rceil$ is said to be {\rm locally observable 
at} ${\bm x}_0$ if rank~${\cal O}_X ({\bm x}_0) = d$ and {\rm globally 
observable} if rank~${\cal O}_X ({\bm x}) = d$ for every 
${\bm x} \in {\cal M}$.\cite{Her77}
\end{defi}

Since for large dimensional systems the observability matrix ${\cal O}_X$ 
is analytically intractable,\cite{Bia15} a procedure was developed to compute 
the symbolic observability matrix and its determinant as 
follows.\cite{Let09,Bia15,Let18} 
The Jacobian matrix of the system under study is transformed into a symbolic 
Jacobian matrix ${\cal J}$ whose elements ${ J}_{lk}$, are 1 when $J_{lk}$ is
constant ($l,k=1,\dots,d$), $\1$ when $J_{lk}$ is polynomial, and $\2$ when 
it is rational. From the symbolic Jacobian matrix, a symbolic observability 
matrix $\tilde{\cal O}_X$ is constructed using an algebra working on the symbols
0, 1, $\1$ and $\2$. The number $N_1$, $N_{\1}$, and $N_{\2}$ of the
different terms (constant, polynomial and rational) occurring in the symbolic
expression of the determinant Det$\tilde{\cal O}_X$ computed using this algebra 
are counted. The symbolic observability coefficients $\eta_X$ are then computed 
according to the formula\cite{Let18}
\begin{equation}
  \label{defsymcoef}
	\eta_X = 
	\frac{N_1}{N_{\rm tot}} +
	\frac{N_{\1}}{N_{\rm tot}^2} +
	\frac{N_{\2}}{N_{\rm tot}^3} 
\end{equation}
where $N_{\rm tot} = N_1 + N_{\1} + N_{\2}$.  The values of $\eta_X$ range from 0 to 1, with $\eta_X=1$ indicating that the pair $\lceil \mb{f}, \mb{X} \rceil$ provides a global observability of the system, good observability if $\eta_X>0.75$ and poor otherwise.\cite{Sen16} 

Let us illustrate this with the R\"ossler system \cite{Ros76c}
\begin{equation}
	\label{ros76}
	\left\{
		\begin{array}{l}
			\dot{x} = -y -z \\[0.1cm]
			\dot{y} = x + a y \\[0.1cm]
			\dot{z} = b + z(x-c) \,  
		\end{array}
	\right.
\end{equation}
whose Jacobian matrix is
\begin{equation}\label{Jacobian}
\cal J = 
  \left[
    \begin{array}{ccc}
      0 & -1 & -1 \\
      1 & a & 0 \\
      z & 0 & x -c
     \end{array}
    \right]\, .
\end{equation}
Let us assume that the measurement function is $ h({\bm x}) = x$, namely 
${\bm X}=\left[ \begin{array}{ccc} x & \dot{x} & \ddot{x} \end{array} 
\right]$. Then the observability matrix is
\begin{equation}
  {\cal O}_{x^3} = 
  \left[
	\begin{array}{ccc}
		1 & 0 & 0 \\[0.1cm]
		0 & -1 & -1 \\[0.1cm]
		-1-z & -a & c-x \\[0.1cm]
	\end{array}
  \right]
\end{equation}
where the subscript $x^3$ designates the first three Lie derivative of $x$, 
that is, $x$, $\dot{x}$ and $\ddot{x}$. The determinant of this observability
matrix is
\begin{equation}
  \mbox{Det } {\cal O}_{x^3} ({\bm x}) =
  1 \cdot \left( 1 \cdot (c-x) - a \cdot 1 \right) = x -a -c \, ,
\end{equation}
which is null in the singular observability manifold defined as\cite{Fru12}
\begin{equation}
  {\cal M}^{\rm obs}_{x^3} = 
	\left\{ \displaystyle {\bm x} \in {\cal M} \subset \mathbb{R}^3
	~|~ x = a + c \right\} \, . 
      \end{equation}
The system (\ref{system}) is not observable from ${\cal M}^{\rm obs}_{x^3}$ 
when $x$ is measured. To answer how good is the 
observability of the pair $\lceil \mb{f}, \mb{X} \rceil$, we resort to the 
symbolic approach described above. The symbolic representation of 
(\ref{Jacobian}) is
\begin{equation}
  \label{syjaros}
  \tilde{\cal J} = 
  \left[
    \begin{array}{ccc}
      0 & 1 & 1 \\
      1 & 1 & 0 \\
      \1 & 0 & \1 
    \end{array}
  \right] 
\end{equation}
and the corresponding symbolic observability matrix is
\begin{equation}
	\tilde{\cal O}_{x^3} =
  \left[
        \begin{array}{ccc}
                1 & 0 & 0 \\[0.1cm]
                0 & 1 & 1 \\[0.1cm]
                \1 & 1 & \1 \\[0.1cm]
        \end{array}
  \right]\, , 
\end{equation}
as detailed in Ref. \onlinecite{Let18}.
We can compute the determinant of $\tilde{\cal O}$ using the symbolic rules 
described in Ref.~\cite{Let18} for the product and the sum
of symbolic terms, leading to
\begin{equation}
	\mbox{Det } \tilde{\cal O}_{x^3} ({\bm x}) =
  1 \otimes \left( 1 \otimes \1 \oplus 1 \otimes 1 \right) \, . 
\end{equation}
Notice that there is no subtraction in the symbolic algebra.\cite{Bia15}
There are $N_1 = 4$ constant terms and $N_{\1} = 1$ rational one. According to
Eq. (\ref{defsymcoef}), the symbolic observability coefficient is thus
\begin{equation}
  \eta_{x^3} = \frac{4}{5} + \frac{1}{5^2} = 0.84 \, . 
\end{equation}
Proceeding in a similar way, the two other symbolic observability coefficients 
are $\eta_{y^3} = 1.00$, and $\eta_{z^3} = 0.44$. Variable $y$ therefore 
provides a global observability of the pair $\lceil \mb{f}, \mb{X} \rceil$ 
where 
${\bm X}=\left[ \begin{array}{ccc} y & \dot{y} & \ddot{y} \end{array} \right]$.
The variable $y$ is thus the preferred measured variable for constructing an 
observer of the R\"ossler dynamics. 

\begin{figure}[ht]
  \centering
  \includegraphics[width=0.15\textwidth]{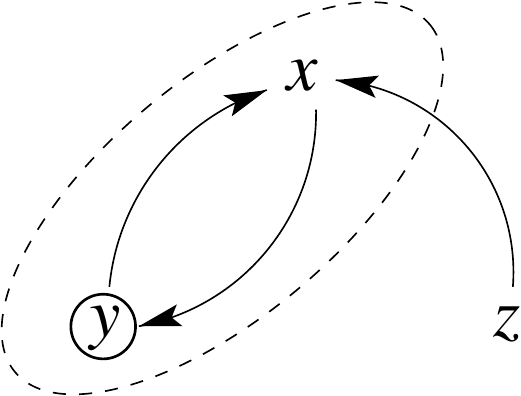} \\[-0.3cm]
  \caption{Pruned fluence graph of the R\"ossler system where an edge is
drawn between $i$ and $j$ nodes (variables $x$, $y$, and $z$) whenever
$J_{ij}$ in Eq.~(\ref{syjaros}) is a nonzero
constant. A dashed oval surrounds the root strongly connected component
(rSCC).  Edges $i \rightarrow i$ (self-loops) are omitted since they do not contribute
to the determination of the rSCC. Variable $y$ is encircled as it
provides global observability of the system. }
  \label{pfgros}
\end{figure}

From the symbolic Jacobian matrix (\ref{syjaros}), we can get a first graphical 
selection of the variables eventually rendering the system globally observable. 
It is based on the concept of {\it root strongly connected components} 
(rSCC) applied to a pruned fluence graph encoding the
interdependence (links) between the state variables
(nodes).\cite{Liu13} A rSCC is the largest subgraph in which there is
a directed path from every node to every other node and with no links
going out: therefore, any node in the rSCC has information about the others and the 
sensor can be placed in any of them. As described in Ref.~\onlinecite{Let18b}, 
the pruned fluence graph is constructed retaining only the nonzero constant 
elements ${J}_{lk}$, that is, nonlinear terms are disregarded as they diminish 
the observability (specifically the terms $J_{31}$  and $J_{33}$ of the R\"ossler system). 
Self-loops are also not taken into account since they do
not contribute to the rSCC. An example is shown in Fig.\ \ref{pfgros} for the R\"ossler system. It has a single rSCC (dashed oval) containing the variables $x$ 
and $y$. The variable $z$ cannot provide a global observability of the 
R\"ossler system, a result which can be analytically proved by computing the 
determinant Det~${\cal O}_{z^3} = -z^2$ which vanishes for $z=0$: there is
a non-empty singular observability manifold.

This graphical approach can be used to determine the observability of networks 
of coupled dynamical systems by using the condition that a sensor should be 
placed at one variable in each rSCC. In the next sections, we will show how 
this graphical approach can be applied to networks of $N$ diffusively coupled
R\"ossler oscillators. 

\section{Observability of coupled dynamical systems}\label{observability}

Let us consider a network of $N$ diffusively coupled R\"ossler oscillators, 
each of them being governed by the vector field
\begin{equation}
  {\bm f}_i = (-y_i-z_i,x_i+a_iy_i,b+z_i(x_i-c)) 
\end{equation}
where $a_i$ is the parameter used to account for the network heterogeneity.
The $i$th node dynamics  is thus governed by
\begin{equation}
  \label{rosnet}
  \dot{\bm x}_i 
   = {\bm f}_i(a_i,{\bm x}_i) 
   + \rho \sum_j {A}_{ij}
	\left[ \displaystyle {\bm g}({\bm x}_j)-{\bm g}({\bm x}_i) \right] \, ,
\end{equation}
for $i=1,\dots,N$, where $\rho$ is the coupling strength, $A=(A_{ij})$ are the 
entries of the adjacency matrix, describing whether nodes $i$ and $j$ are 
coupled if $A_{ij}=1$ or not ($A_{ij}=0$), and ${\bm g}: \mathbb{R}^d \mapsto
\mathbb{R}^d$ is the coupling function. We will consider the R\"ossler 
oscillators coupled through any of the three
variables, that is, ${\bm g}_x(\bm x)=(x,0,0)$, ${\bm g}_y(\bm
x)=(0,y,0)$, or ${\bm g}_z(\bm x)=(0,0,z)$. 

\subsection{Observability of dyads of R\"ossler systems}

Let us now investigate the placement of sensors in two R\"ossler systems 
unidirectionally coupled ($A_{12}=0$ and $A_{21}=1$) using the graphical
approach described in the Section \ref{obserap}. 
The symbolic Jacobian matrix of the dyad is:
\begin{equation}
  \tilde{\cal J} =
  \left[
    \begin{array}{cccccc}
      0 & 1 & 1 & 0 & 0 & 0 \\[0.1cm]
      1 & 1 & 0 & 0 & 0 & 0 \\[0.1cm]
      \1 & 0 & \1 & 0 & 0 & 0 \\[0.1cm]
	    {\bf 1} & 0 & 0& 0 & 1 & 1 \\[0.1cm]
	    0 & {\bf 1} & 0& 1 & 1 & 0 \\[0.1cm]
	    0 & 0 & {\bf 1} & \1 & 0 & \1 \\[0.1cm]
    \end{array}
  \right]
\end{equation}
where the $3\times 3$  diagonal blocks coincide with Eq.~(\ref{syjaros}) for an 
isolated R\"ossler system and off-diagonal entries ${J}_{ij}$ in bold 
represent the unidirectional coupling from node 1 to node 2 through each one of 
the three variables simultaneously. In the following, we will consider the three coupling 
schemes, ${\bm g}_x$, ${\bm g}_y$, and ${\bm g}_z$, separately.

\begin{figure}[ht]
  \centering
  \begin{tabular}{ccccc}
    \includegraphics[width=0.24\textwidth]{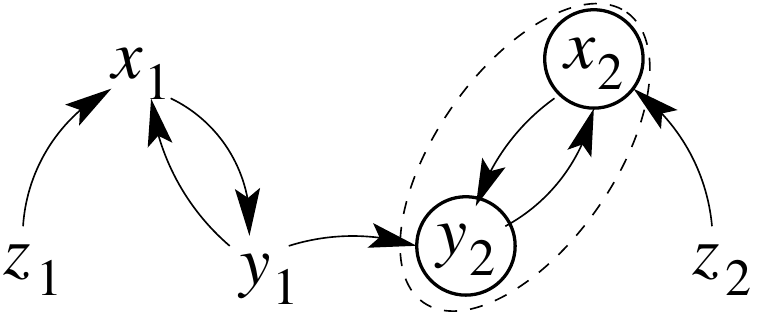} \\
    (a) Dyad with ${\bm g}_y$ coupling: 1 rSCC \\[0.2cm]
    \includegraphics[width=0.26\textwidth]{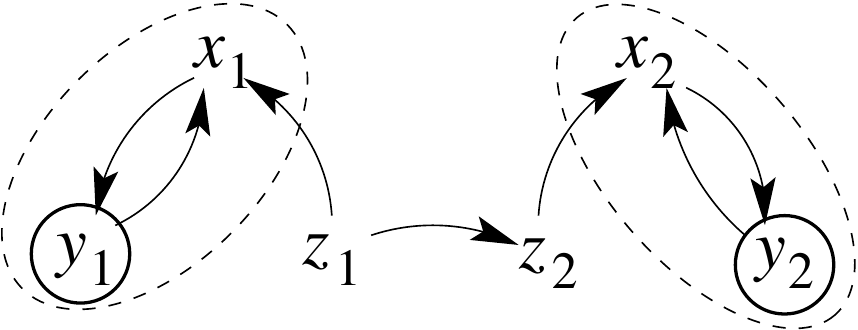} \\
    (b) Dyad with ${\bm g}_z$ coupling: 2 rSCC \\[0.2cm]
    \includegraphics[width=0.28\textwidth]{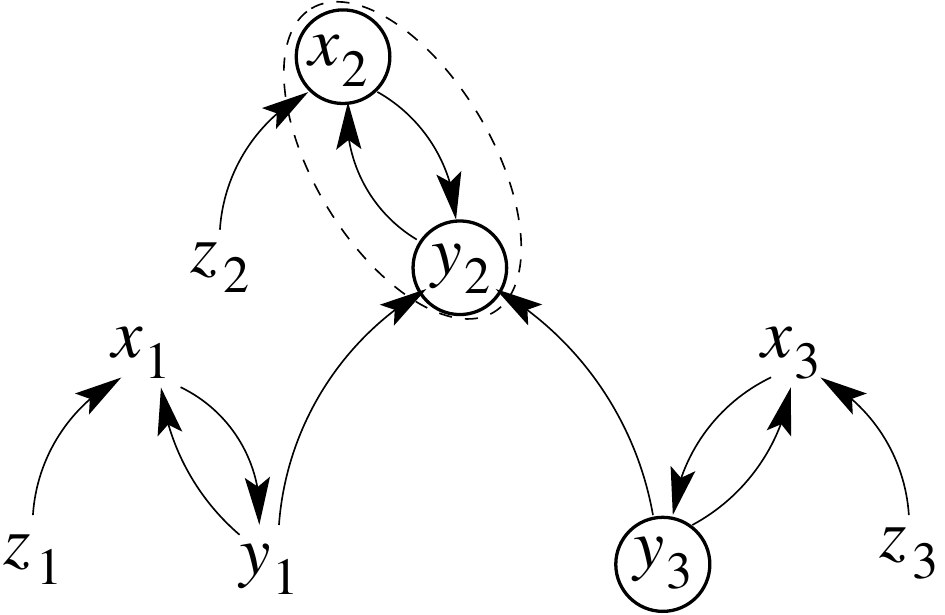} \\
    (c) Triad with ${\bm g}_y$ coupling: 1 rSCC \\[-0.2cm]
  \end{tabular}
  \caption{Pruned fluence graphs of two dyads made of R\"ossler systems coupled 
by different variables and of a triad of them. The root strongly connected 
	components (rSCC) are shown in dashed lines. Sensor variables are encircled.} 
  \label{dyadgra}
\end{figure}

The pruned fluence graphs constructed for ${\bm g}_y$ and ${\bm g}_z$ are shown 
in Fig.\ \ref{dyadgra}(a) and \ref{dyadgra}(b), respectively. The corresponding 
rSCC are marked with oval dashed lines. There is a 
single rSCC when the two R\"ossler systems are coupled through variable $y$: it
contains variables $x_2$ and $y_2$ (the rSCC is the same for the 
${\bm g}_x$ coupling). There are two rSCCs when the R\"ossler systems are 
coupled through variable $z$.

Therefore, a dyad of R\"ossler systems unidirectionally coupled through the 
variable $y$ (or $x$) is potentially globally observable by just measuring 
variable $x_2$ and/or $y_2$ [encircled variables in Fig.~\ref{dyadgra}(a)]. To 
answer this in a more accurate way, we applied a systematic computation of the 
symbolic observability coefficients\cite{Sen19} and found that, when the 
reconstructed space is spanned by the vector
\begin{equation}
  \label{dyadx2y2}
  \mb{X} = 
  \left[ 
    \begin{array}{cccccc}  
      x_2 & \dot{x}_2 & y_2 & \dot{y}_2 & \ddot{y}_2 & \stackrel{...}{y}_2 
    \end{array}
  \right]^{\T}  \, ,
\end{equation}
$\eta_{x_2^2 y_2^4}=1$, suggesting that the pair $\lceil \mb{f}_1 + \mb{f}_2,
x_2^2 y_2^4 \rceil$ is globally observable. This is analytically checked with
the determinant Det~${\cal O}_{x_2^2 y_2^4} = - \rho^3$, confirming the global
observability of the R\"ossler system by measuring variables $x_2$ and
$y_2$ (until $\rho \neq 0$). There are two other combinations providing a 
global observability: Det~${\cal O}_{x_2^5 y_2} = - \rho^3$ (measuring $x_2$ 
and $y_2$) and Det~${\cal O}_{x_2^5 z_2} = \rho^3$ (measuring $x_2$ and $z_2$). 
Finally, there is another combination providing a good observability 
($\eta_{x_1^5z_2} = 0.91$) with the reconstructed vector
\begin{equation}
  \label{dyady2z2}
   \mb{X} = 
  \left[ 
    \begin{array}{cccccc}  
      y_2 & \dot{y}_2 & \ddot{y}_2 & \stackrel{...}{y}_2 &
	    \stackrel{....}{y}_2 & z_2
    \end{array}
  \right]^{\T}  \, :
\end{equation}
the corresponding analytical determinant 
$\mbox{Det }{\cal O}_{y_2^5z_2} = \rho^3 (c + a_1 -x_1) $ is a first-degree
polynomial (as expected for $\eta_X>0.75$).\cite{Sen16} There is a singular 
observability manifold ${\cal M}^{\rm obs}_{y_2^5 z_2} \equiv
  \left\{ \displaystyle \mb{x} \in {\cal M} ~|~ x_1 = c + a_1 \right\}$
which cannot be observed from their measurements. Such manifold should degrade
the performance of any observer built from variables $y_2$ and $z_2$.

When the coupling ${\bm g}_z$ is chosen, the pruned fluence graph in Fig.\
\ref{dyadgra}(c) displays two rSCCs involving the two nodes and, therefore, one
variable has to be chosen in each of them. In that case, the most obvious 
choice is to measure variable $y$ in each node. According to this analysis, the 
coupling function has a profound impact in the observability of a network: 
while couplings ${\bm g}_x$ or ${\bm g}_y$ permits a pair of coupled R\"osslers 
to be globally observable by just measuring in one of them, it forces to 
perform measures in both systems when they are coupled through the $z$ 
variable. 

\subsection{Observability of triads and larger networks}\label{sectriadnet}

From the study of a simple dyad, we have shown that the observability
and, consequently, the set of variables to be measured, depends
largely  on the coupling function. Therefore, let us now consider a triad of
R\"ossler systems coupled as shown in Fig.~\ref{dyadgra}(c) using the variable 
$y$ as the coupling variable connecting nodes 1 and 3 to node 2. A graphical 
approach returns a single rSCC containing variables $x_2$ and $y_2$. However, a 
systematic study of all possible reconstructed vectors based on 
these two variables shows that it is not possible to get a global observability 
of more than two nodes from measurements on a single one.\cite{Sen19} A global 
observability of our triad is therefore obtained by measuring variables 
$x_2$ and $y_2$ in node 2, and variable $y_3$ in node 3 [encircled variables
in Fig.~\ref{dyadgra}(c)]. Using these three variables and their derivatives, 
it is possible to get a reconstructed vector ${\bm X}$ providing a global 
observability since $\eta_{x_2^2 y_2^4 y_3^3} = 1$. Notice that we also have 
$\eta_{y_3^3 x_2^2 y_2^4} = 1$ if the roles of nodes 1 and 3 are exchanged.

From the knowledge gathered for dyads and triads of systems like the R\"ossler 
one, we can infer the following propositions for larger networks of any 
dynamical system.\cite{Sen19}

\begin{prop}
  \label{propfully}
When the node dynamics is globally observable from one of its variables, then a 
dynamical network is globally observable if that variable is measured at each
node ($m=N$), independently from the coupling function and topology, even when
the network is not completely connected.
\end{prop}

\begin{coro}
When the number $N_m$ of measured nodes is such that $N_m  < N$, by definition,
the choice of the variables to measure not only depends on the adjacency
matrix $A$ and coupling function ${\bm g}$, but also, on the node dynamics.
\end{coro}

\begin{prop}
  \label{propz}
When a network of $N$ R\"ossler systems is coupled by the variable $z$,
then $N_m=N$ nodes must be measured to obtain a global observability.
\end{prop}

\begin{prop}
  \label{prop1}
In a network of $N$ R\"ossler systems linearly coupled, it is not possible to 
reconstruct with a global observability the space associated with three nodes 
from measurements in a single node.
\end{prop}

\begin{coro}
  \label{coromin}
A global observability of a network of $N$ R\"ossler systems is obtained from
measurements of at least $m=N$ variables in at least $N_m=\frac{N}{2}$ nodes. 
\end{coro}

Therefore, in order to investigate the observability of a network of dynamical
systems and address the problem of the choice of a set of sensors,
we first need to tackle the observability at the node level, 
that is, of the nodal dynamics, and then to proceed with the
observability of a pair of nodes to incorporate the role of
the coupling function. 

In the next sections, we will explain the strategy to select the $N_m$ nodes 
and how we can build from them an observer for a whole network of R\"ossler 
oscillators.

\section{Observers}
\label{observer}

\subsection{Observer of a single R\"ossler system}

An observer can be easily constructed by inverting the coordinate 
transformation $\Phi : \mb{x} \mapsto \mb{X}$ between the original state space
$\mathbb{R}^3 (\mb{x})$ and the reconstructed space $\mathbb{R}^3 (\mb{X})$. 
When the system is globally observable, that is, when the determinant of the 
Jacobian matrix of the map $\Phi$ never vanishes, the map is easily inverted 
and the differentiator ${\rm D}=\Phi^{-1}$ is defined over the whole state 
space.\cite{Her77}

From the observability analysis of the R\"ossler system performed in
Sec.~\ref{obserap}, we will construct observers using the differentiators 
obtained by inverting the three maps investigated. Let us start with 
${\bm X}=\left[ \begin{array}{ccc} y & \dot{y} & \ddot{y} \end{array} \right]$, 
which provides a global observability. In this case case, the differentiator 
reads
\begin{equation}
  \label{Dy}
	D_y \equiv
  \Phi^{-1}_{y^3} = 
  \left|
    \begin{array}{l}
      \displaystyle
	    \tilde{x} = - a X_1 + X_2 \\[0.1cm]
      \displaystyle
      y = X_1 \\[0.1cm]
      \displaystyle
	    \tilde{z} = -X_1 + a X_2 - X_3 \,  
    \end{array}
  \right.
\end{equation}
where $X_1=y$, $X_2 = \dot{y}$, and $X_3 = \ddot{y}$, and $\tilde{x}$ and 
$\tilde{z}$ are the estimated variables. An alternative observer can be 
constructed from the governing equations (\ref{ros76}), and replacing the 
third equation, obtaining: 
\begin{equation}
  \label{Oy}
	O_y \equiv
  \left|
    \begin{array}{l}
      \displaystyle
	    \tilde{x} = - a X_1 + X_2 \\[0.1cm]
      \displaystyle
      y = X_1 \\[0.1cm]
    \tilde{z} = -X_1 - \dot{\tilde x} \, . 
    \end{array}
  \right.
\end{equation}
where the derivative of $\tilde x$ has to be computed from the estimated 
variable $\tilde{x}$.

When variable $x$ is measured, that is, $X_1 = x$, the differentiator
reads
\begin{equation}
	D_x \equiv
  \Phi^{-1}_{x^3} = 
  \left|
    \begin{array}{l}
      x = X_1 \\[0.1cm]
      \displaystyle
	    \tilde{y} = \frac{b + X_1 + c X_2 + X_3 -X_1 X_2}{X_1 -a -c} \\[0.3cm]
      \displaystyle
	    \tilde{z} = \frac{-b  -X_1 + a X_2 - X_3}{X_1 - a -c} \,  
    \end{array}
  \right.\label{Dx}
\end{equation}
where $X_2 = \dot{x}$, and $X_3 = \ddot{x}$. This differentiator is not defined
in the singular observability manifold \cite{Fru12}
\begin{equation}
	{\cal M}^{\rm obs}_x
	\equiv
	\left\{ \displaystyle (x,y,z) \in \mathbb{R}^3 ~|~ x = a+c \right\}
\end{equation}
that is visited by the chaotic trajectory with a null Lebesgue measure. Despite 
the good observability offered by this variable ($\eta_x^3=0.84$), the 
performance of the observer is actually compromised due to the singular 
observability manifold.

Finally, when variable $z$ is measured, $X_1 = z$, a differentiator 
reads
\begin{equation}
	D_z \equiv
  \Phi^{-1}_{z^3} = 
  \left|
    \begin{array}{l}
      \displaystyle
      \tilde{x} = \frac{-b + c X_1 + X_2}{X_1} \\[0.3cm]
      \displaystyle
      \tilde{y} = - X_1 + \frac{-b X_2 - X_1X_3 + X_2^2}{X_1^2} \\[0.3cm] 
      z = X_1 \, 
    \end{array}
  \right.\label{Dz}
\end{equation}
where $X_2 = \dot{z}$, and $X_3 = \ddot{z}$. This differentiator is not defined
in the singular observability manifold 
\begin{equation}
	{\cal M}^{\rm obs}_z
	\equiv
	\left\{ \displaystyle (x,y,z) \in \mathbb{R}^3 ~|~ z = 0 \right\}
\end{equation}
that is visited by the chaotic trajectory with a null Lebesgue measure. There 
is no possibility to easily construct an observer from the R\"ossler equations
(\ref{ros76}) when variable $z$ is measured.

In order to quantify the goodness of these observers, we compute the
normalized root-mean-square error (NRMSE) of the estimated state variable 
$ \tilde{\bm x}=({\tilde x}(t_n), {\tilde y}(t_n), {\tilde z}(t_n))$ of 
${\bm x}(t_n)$ at time $t_n$ in a time window of length $T=t_M-t_1$ with 
$n=1,\dots,M$,
\begin{equation}
  \epsilon 
  = \frac{1}{3}\left(\epsilon_{\tilde x}
	+\epsilon_{\tilde y} + \epsilon_{\tilde z}\right)  \label{err_amp}
\end{equation}
being $\epsilon_{\tilde x}$ the NRMSE of the $\tilde x$ variable and computed 
as
\begin{equation}
  \epsilon_{\tilde x}   =
  \frac{1}{\Delta_x}
	\sqrt{\frac{1}{M} \displaystyle \sum_{n=1}^{M} 
	\left[ \displaystyle \tilde{x}(t_n)-x(t_n) \right]^2}
\end{equation}
where $\Delta_x = x_{\rm max} - x_{\rm min}$ is the normalization
factor. Similar errors are defined for the other variables.

It is also interesting to monitor how closely an observer is able to
capture the phase of the dynamics.
We compute the phase using a linear
interpolation between two intersections with a Poincar\'e section\cite{Osi03}
defined as 
\begin{equation}
  {\cal P} \equiv \left\{
   ( x_n, z_n) \in \mathbb{R}^2 ~|~ y_n = y_{\rm p}, \dot{y}_n < 0 \right\}
\end{equation}
where 
\[ y_{\rm p} = - \frac{c - \sqrt{c^2 - 4 ab}}{2a} \]
and $(x_n, y_n, z_n)$ are the coordinates of the $n$th intersection of the 
trajectory with the Poincar\'e section. Then, as for the NRMSE, we 
compute the normalized error of the estimated phase $\tilde \phi$ as
\begin{equation}
  \epsilon_{\tilde \phi} = \frac{1}{2\pi}
  \sqrt{\displaystyle \frac{1}{M} 
  \sum_{n=1}^{M} \left| \displaystyle \tilde{\phi}(t_n)-\phi(t_n) \right|}
	\, . \label{err_ph}
\end{equation}

System (\ref{ros76}) was integrated using a 4th-order Runge-Kutta with a time 
step ${\rm d} t=0.01$. For any of the state observers, derivatives of the
measured variables at time $t_k$ (in terms of d$t$) are computed using the 
first-order finite difference scheme as
\begin{equation}
  \label{fidische}
  \left\{
    \begin{array}{l}
	    \displaystyle
      \dot{v}(t_{k-1}) = \frac{v(t_k) - v(t_{k-2})}{2 \, {\rm d}t} \\[0.3cm]
	    \displaystyle
      \ddot{v}(t_{k-1}) 
	    = \frac{v(t_{k}) - 2 v(t_{k-1}) + v(t_{k-2})}{{\rm d}t^2} \, ,
    \end{array}
  \right.
\end{equation}
with $v=x,y,z$. Such a scheme implies that the derivatives are known at the 
discrete times $(k-1) {\rm d}t$ and $(k-2){\rm d}t$. The differentiator returns the state at
$(k-1) \, {\rm d}t$ while the observer $O_y$ returns it a 
$(k-2) \, {\rm d}t$ since the derivative of $\tilde{x}$ requires an additional
time step.

Table~\ref{errortable} summarizes the normalized errors (in percentage) for the 
amplitude and phase of the estimated variables using the differentiators and 
observers from Eqs.~(\ref{Dy}), (\ref{Oy}), (\ref{Dx}), and (\ref{Dz}) for the
R\"ossler system with $a=0.43$ (other parameter values lead to similar errors) 
and a time window of $\tau = 600$ time units. As expected, the smallest errors 
are obtained with the observers using the variable $y$ which provides global 
observability. 

\begin{table}[ht]
	\centering
	\caption{Normalized errors of the variable estimation $\epsilon_{\tilde
          x}$ and of the phase $\epsilon_{\tilde \phi}$ obtained with
        the  differentiators and observators constructed from the
        R\"ossler equations with $a=0.432$. }
	\label{errobs}
	\begin{tabular}{cccccccc}
		\\[-0.3cm]
		\hline \hline
		\\[-0.3cm]
		& D$_{x}$ & D$_{y}$ & D$_{z}$ & O$_{y}$ & 
		\\[0.1cm]
		\hline
		\\[-0.3cm]
         $\epsilon:$ & 0.18\% & 0.0004\% & 0.0015\% & 0.001\% \\[0.1cm]
$\epsilon_{\tilde\phi}:$ & 0.1427\% & 0.0000\% & 0.0737\% & 0.0000\% \\[0.1cm]
		\hline \hline
	\end{tabular}\label{errortable}
\end{table}

\begin{figure}[h]
  \centering
  \includegraphics[width=0.34\textwidth]{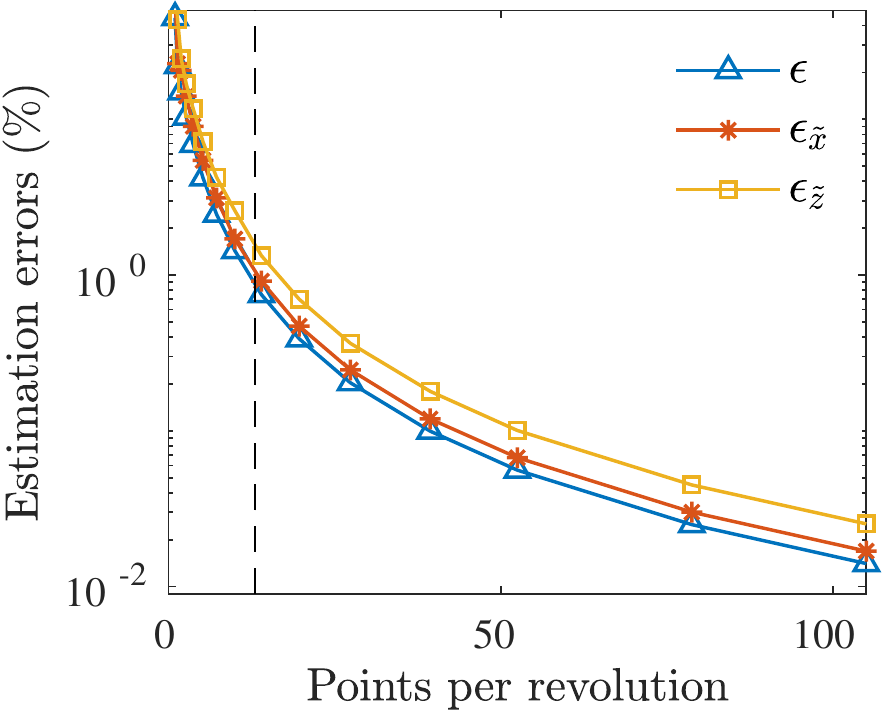} \\[-0.3cm]
  \caption{Estimation errors --- $\epsilon_{\tilde{x}}$ and 
$\epsilon_{\tilde{z}}$ --- as a function of the number of points 
per revolution. The Nyquist criterion returns at least 13 points per revolution 
	to avoid undersampling effect.}
  \label{errNr}
\end{figure}

It is well known that the results of the observer depends on the sampling 
frequency of the measured variable.\cite{Nyq24,Let22a} Typically, the sampling 
frequency $f_{\rm s}$ has to be such that $2 f_{\rm s} > f_{\rm max}$ where 
$f_{\rm max}$ is the highest frequency of the measured variable (here 
$f_{\rm max} \approx 1$~Hz). Since the pseudo-period of the R\"ossler dynamics 
is 6.4~s, at least 13 points per revolution are needed to fulfill the Nyquist 
criterion. Setting a sampling frequency slightly greater than the Nyquist 
frequency $f_{\rm max}$ is sufficient to recover a reasonable estimation of the 
phase, but recovering correctly the variable $x$ and the variable $z$ requires 
at about 50 points per revolution as shown in Fig.\ \ref{errNr} (a common 
requirement for investigating accurately a chaotic dynamics\cite{Fre09b}).

\subsection{Observer of a dyad}

Let us now move to the case of designing a state observer of a dyad of 
R\"ossler systems coupled through variable $y$ as sketched in 
Fig.~\ref{dyadgra}(a) where nodes 1 and 2 are unidirectionally coupled as 
$1 \rightarrowtail 2$. The observability analysis showed that the 6-dimensional 
state space $(x_1,y_1,z_1,x_2,y_2,z_2)$ associated with such a dyad is globally
observable by only measuring variables $x_2$ and $y_2$ of node 2, that is, an 
observer can be constructed without
singular observability manifold. Combining the governing equations for node 
2 and a differentiator built from the reconstructed variable $\tilde{y}_1$
for node 1, we propose the following state observer
$(\tilde{x}_1,\tilde{y}_1,\tilde{z}_1,x_2,y_2,\tilde{z}_2)$ for $\rho>0$:
%
\begin{equation}
  \label{obsdya}
  \left|
    \begin{array}{l}
	    \displaystyle
	    \tilde{x}_1 = -a_1 \tilde{y}_1 + \dot{\tilde{y}}_1 \\[0.1cm]
	    \displaystyle
      \tilde{y}_1 = y_2 + \frac{1}{\rho}{[\dot{y}_2 - x_2 + (\rho-a_2)y_2]} \\[0.1cm]
	    \displaystyle
      \tilde{z}_1 = - \tilde{y}_1 + a_1 \dot{\tilde{y}}_1 
	            - \ddot{\tilde{y}}_1 \\[0.1cm]
      x_2 \\[0.1cm]
      y_2 \\[0.1cm]
	    \displaystyle
      \tilde{z}_2 = -\dot{x}_2 - y_2 \\[0.1cm]
    \end{array}
  \right.
\end{equation}
where $a_1$ and $a_2$ are the $a$ parameters of each node. Here, the observer
returns the state at time $(k-1) \, {\rm d}t$.

\begin{figure}
  \centering
  \begin{tabular}{c}
  \includegraphics[width=0.34\textwidth]{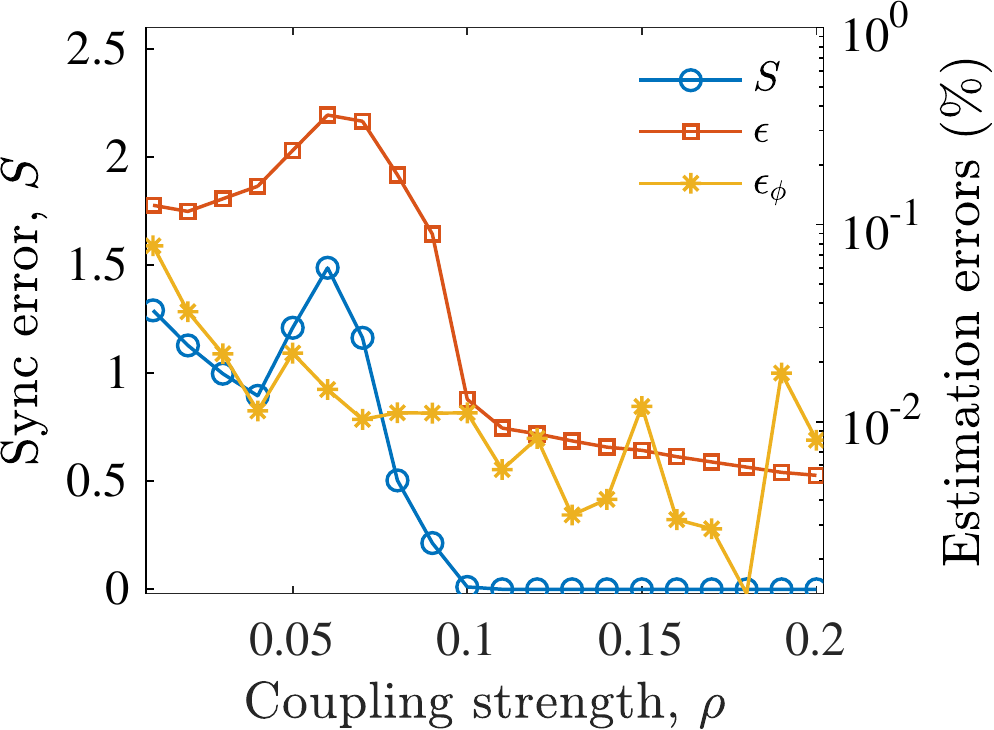} \\
	  (a) Identical nodes \\[0.2cm]
 \includegraphics[width=0.34\textwidth]{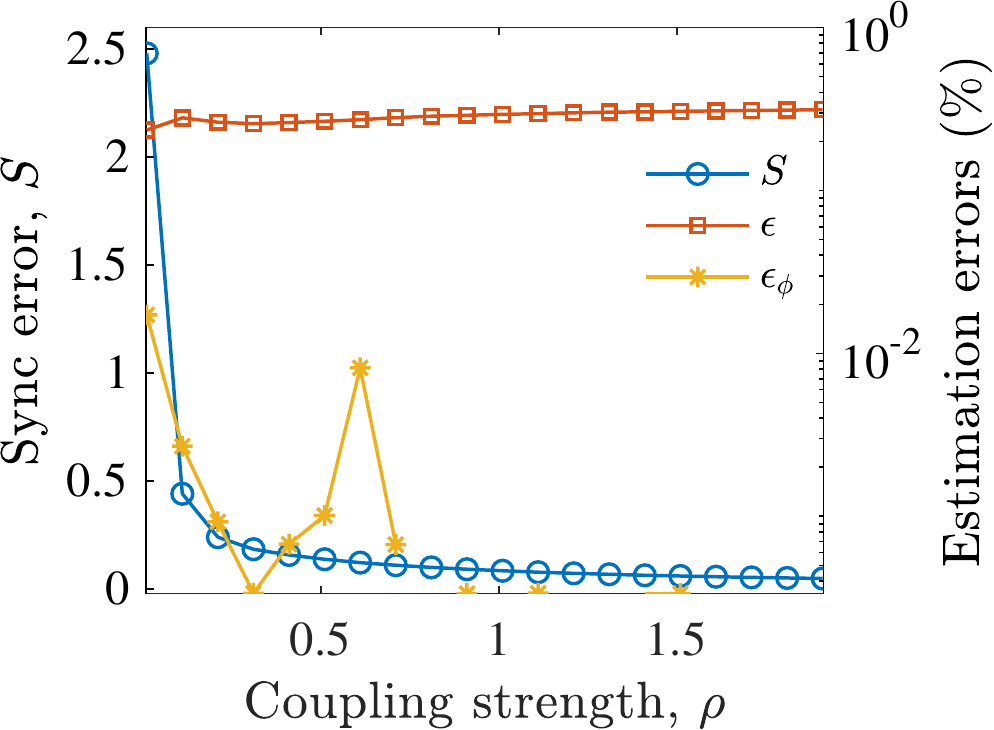} \\
	  (b) Non-identical nodes \\[-0.2cm]
  \end{tabular}
     \caption[]{Dyad observer performance. Synchronization error $S$, 
amplitude error $\epsilon$, and phase reconstruction error $\epsilon_\phi$ as 
a function of the coupling $\rho$ of a dyad with (a) identical 
($a_1=a_2=0.432$) and (b) non identical ($a_1=0.37$, $a_2=0.432$) R\"ossler 
oscillators. The sensor variables are $x_2$ and $y_2$. In each case, the 
synchronization error $S$ is read on the left vertical axis and the estimation
errors in phase ($\epsilon_\phi$) and amplitude ($\epsilon$) are read (in 
percentages) on the right axis. Each point is the average of 10 different 
initial conditions. Other parameter values: $b = 2$ and $c= 4$.}
  \label{figdyad}
\end{figure}

An example of the performance of this observer  as a function of the coupling 
strength $\rho$ is shown in Fig.~\ref{figdyad} for a dyad with two identical 
R\"ossler systems ($a_1=a_2=0.432$) and with two different ones ($a_1=0.37$ and 
$a_2=0.432$). The ensemble average ($N=2$) of the errors in the estimation of 
the amplitude [Eq.~(\ref{err_amp})] and phase [Eq.~(\ref{err_ph})] of the two 
nodes are plotted together with the time averaged synchronization error 
computed as 
\begin{equation}
  S = \frac{1}{M}\frac{2}{N (N-1)}\sum_{n=1}^M  \sum_{\substack{i=1 \\i \ne j}}^N \| \mb{x}_i(t_n)-\mb{x}_j(t_n) \|
\end{equation}
to help us correlating the synchronous state of the dyad and the easiness of 
the prediction: the estimation error of both the amplitude and phase keeps very 
low and stable above a given coupling value even when the nodes are not 
synchronous at all, and when the systems are not identical. 

\section{Network observer}\label{networkobserver}

In Sec.~\ref{sectriadnet}, we concluded that it is not possible to retrieve the 
dynamics of more than two nodes from measurements in a single one and, for 
instance, when we consider the small motif of three nodes shown in 
Fig.~\ref{dyadgra}(c), we have to perform measurements at least in two nodes 
for reconstructing the whole dynamics. The observer for this
configuration needs to measure $x_2$ and $y_2$ to reconstruct the dyad
and $y_3$ to reconstruct the third node. Note that the dyad is not
isolated and one of the sensor variables $y_2$ is receiving input from
$y_3$. Therefore, we propose to generalize the observer in
Eq.~(\ref{obsdya}) as follows:
\begin{equation}
  \label{obsdyadnet}
  \left|
    \begin{array}{l}
            \displaystyle
	    \tilde{x}_i = -a_i \tilde{y}_i + \dot{\tilde{y}}_i \\[0.1cm]
            \displaystyle
      \tilde{y}_i = y_j + \frac{1}{\rho} [\dot{y}_j - x_j -a_j y_j]
	    + \sum_{l=1,l\ne i}^N A_{jl} (y_j -  \tilde{y}_l)] \\[0.1cm]
            \displaystyle
      \tilde{z}_i = - \tilde{y}_i + a_i \dot{\tilde{y}}_i
                    - \ddot{\tilde{y}}_i \\[0.1cm]
      x_j \\[0.1cm]
      y_j \\[0.1cm]
            \displaystyle
      \tilde{z}_j = -\dot{x}_j - y_j \,
      \end{array}
  \right.
\end{equation}
to reconstruct the dyads composed by nodes $i$ and $j$ embedded in
a network (nodes $l$ are neighbors --- according to the adjacency matrix ---
of the sensor $j$ except $i$), and the observer 
\begin{equation}
  \label{obssingle}
  \left|
    \begin{array}{l}  
      	   \displaystyle
      \tilde{x}_k = - a_k y_k + \dot{y}_k \\[0.1cm]
      \displaystyle
      y_k \\[0.1cm]
            \tilde{z}_k = - y_k + a_k \dot{y}_k - \ddot{y}_k  \,  
    \end{array}
  \right.
\end{equation}
for those nodes which have a sensor variable and are not paired. Due to the 
way these observers are constructed, the state (estimated or measured) of the 
nodes connected to the sensor variable of the dyad are needed prior the dyad
reconstruction. Therefore, an iterative procedure is required for networks with 
$N>3$. 

In the following, we will explore first the performance of 
Eqs.~(\ref{obsdyadnet}) and (\ref{obssingle}) for the case of a triad and later 
on we will extend our results to the case of larger networks. 

\subsection{Case of a triad}

\begin{figure}
  \centering
  \begin{tabular}{c}
	  \includegraphics[width=0.34\textwidth]{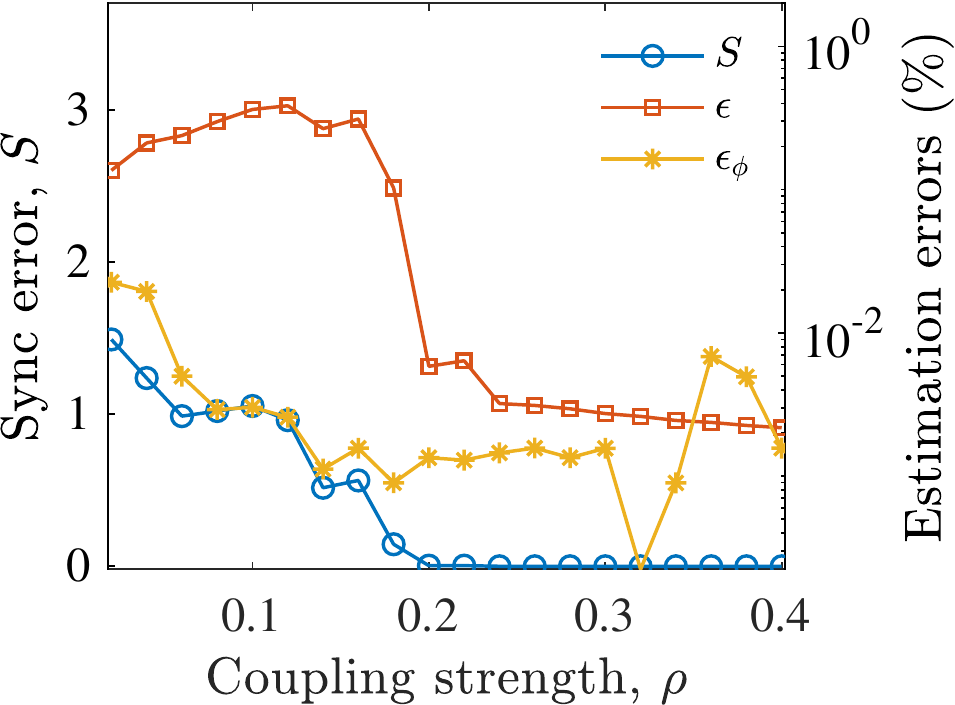} \\ 
	  (a) Identical oscillators \\[0.3cm]
  \includegraphics[width=0.34\textwidth]{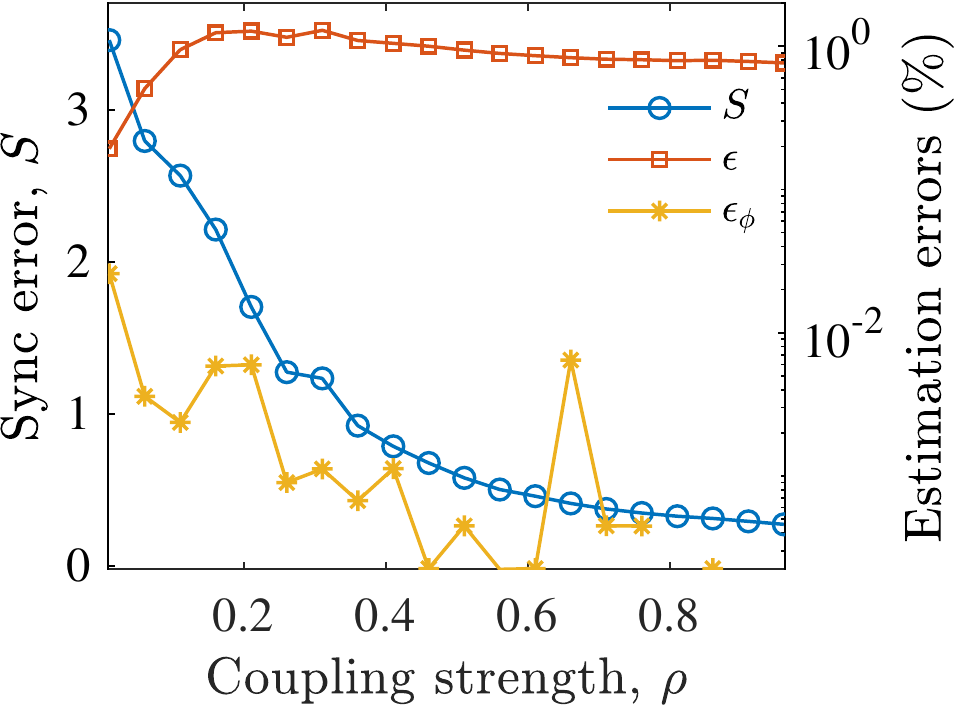} \\
    (b) Non-identical oscillators \\[-0.2cm]
  \end{tabular}
 \caption[]{Triad observer performance. Synchronization error $S$, amplitude 
reconstruction error $\epsilon$, and phase reconstruction error $\epsilon_\phi$
as a function of the coupling $\rho$ of a triad with (a) identical oscillators
($a_1=a_2=a_3=0.432$) and (b) non identical oscillators ($a_1=0.37$, 
$a_2=0.432$, and $a_3=0.52$). In each case, the synchronization error $S$ is 
read on the left vertical axis while the estimation errors ($\epsilon$ and 
$\epsilon_\phi$) are read (in percentages) on the right axis
in log scale. Each point is the average from 10 different initial conditions.}
  \label{figtriad}
\end{figure}

We here consider the case of a triad of R\"ossler systems coupled through 
variable $y$ with the structure $1 \rightarrowtail 2 \leftarrowtail 3$ as shown 
in Fig.~\ref{dyadgra}(c). In Fig.~\ref{figtriad} we show the performance of the 
state observer with identical and non identical oscillators. The behavior is
quite similar to that of a dyad but with slightly higher  amplitude errors
when the triad is composed of non identical systems.

\begin{figure}
  \centering
	\includegraphics[width=0.48\textwidth]{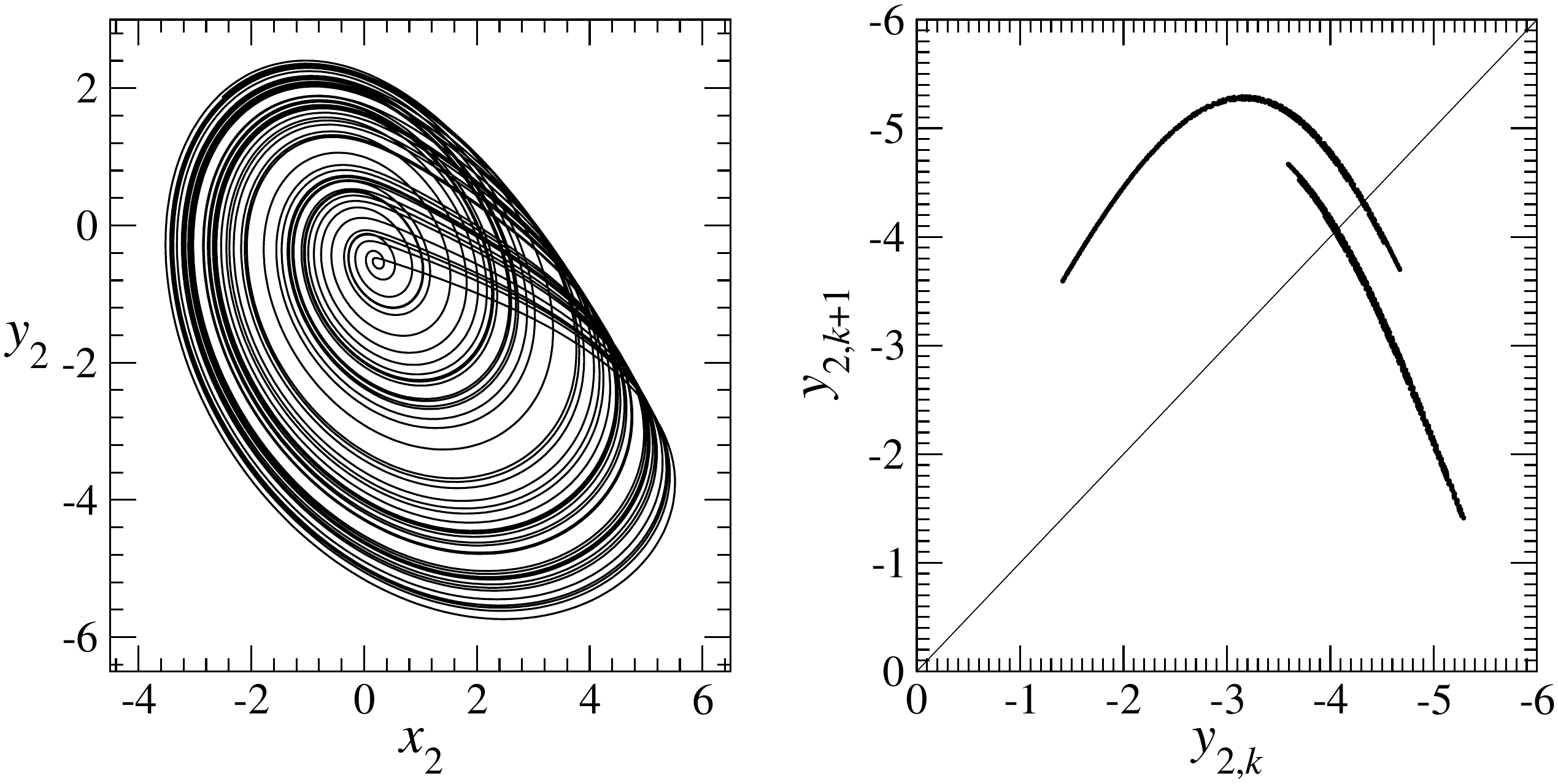} \\
	(a) Dyad with $1 \rightarrowtail 2$ \\[0.2cm]
	\includegraphics[width=0.48\textwidth]{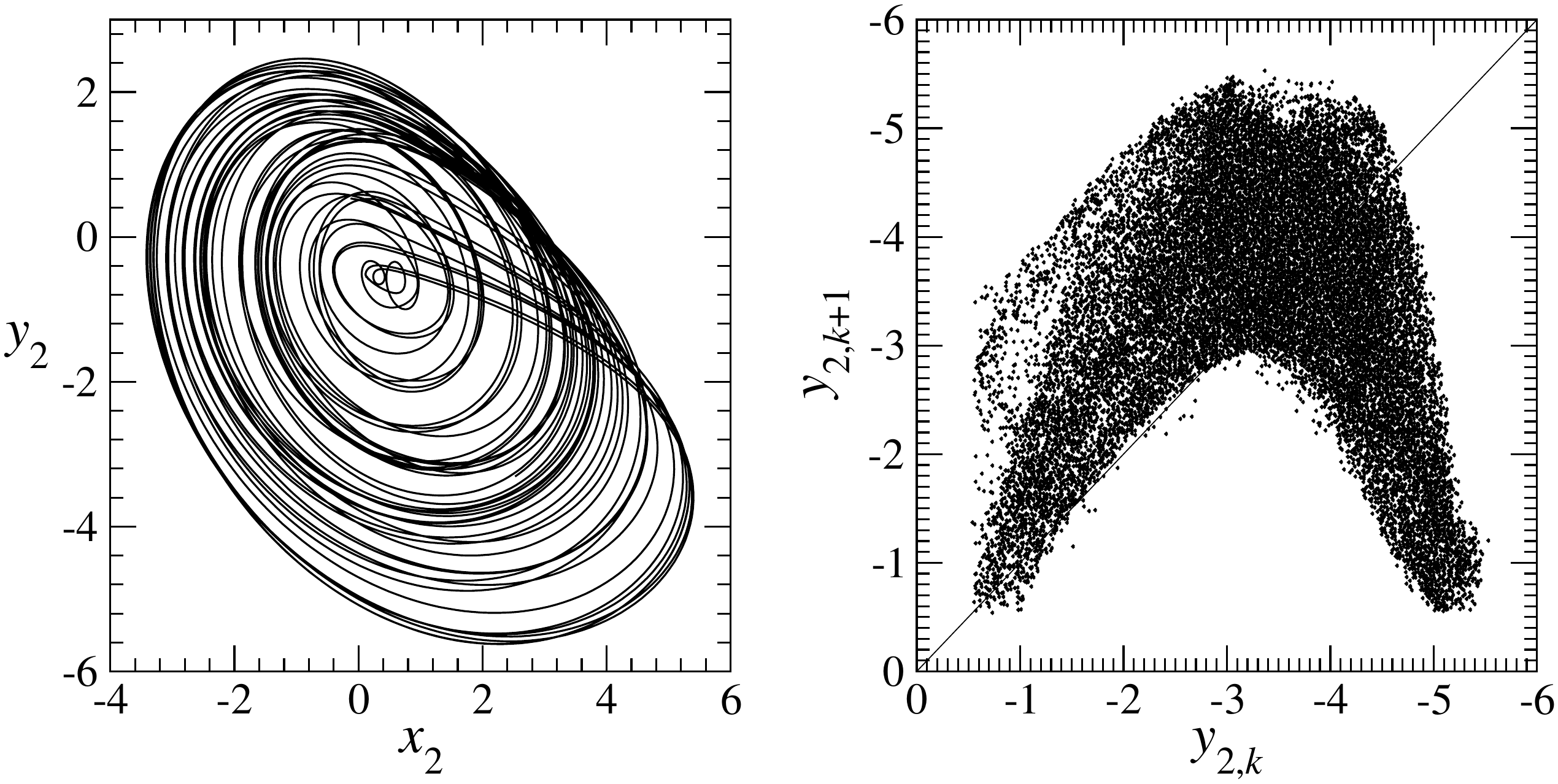} \\
	(b) Triad with $1 \rightarrowtail 2 \leftarrowtail 3$ 
	\\[-0.2cm]
  \caption{Dynamics of the node 2~in the (a) dyad and (b) triad investigated in 
the right panels of Figs.~\ref{figdyad} and \ref{figtriad}, respectively, for 
$\rho=0.1$. State portraits (left panels) and the corresponding first-return 
maps (right panels) are depicted.}
  \label{dydytri}
\end{figure}

Notice that the dynamics of node 2 is now far more complex than in the case of 
the dyad of non identical oscillators: the first-return map of the sensor $y_2$ 
in the triad [Fig.\ \ref{dydytri}(b)] is very thick although the global shape 
still suggests two monotone branches as for the sensor $y_2$ in the dyad
[Fig. \ref{dydytri}(a)]. When the first-return map has a significant thickness 
$\epsilon$, there are more than one periodic orbit associated with a given 
orbital sequence: it is thus possible to have two different linking numbers 
--- the number of times one periodic orbit cycles around the other --- between 
orbits characterized by the same symbolic sequences, respectively. It is said 
that the attractor is $\epsilon$-topologically equivalent to the 
template which could be constructed if the thickness $\epsilon$ was removed 
(see Refs.~\onlinecite{Let95a,Man18b} for details). It appears rather challenging to 
retrieve the dynamics of node 1 (a period-2 limit cycle) from such a complex
dynamics observed in node 2. 

\subsection{Observer hierarchical dependence}

According to Corollary \ref{coromin} and Proposition \ref{prop1}, an observer 
for a network can be constructed from Eqs.~(\ref{obsdyadnet}) and 
(\ref{obssingle}) by assembling the nodes by pairs and designate as sensor one 
member of each pair. There are multiple configurations for decomposing the 
network in pairs but in order to ensure a proper functioning of the algorithm 
and to unravel the coupling term in Eq.~(\ref{obsdyadnet}), we have to proceed 
in a hierarchical way, pairing first the nodes with lower degree; moreover, it
reduces the possibilities and the pairing can be performed automatically.

\begin{figure}
  \vspace{-0.4cm}
  \centering
  \includegraphics[width=0.48\textwidth]{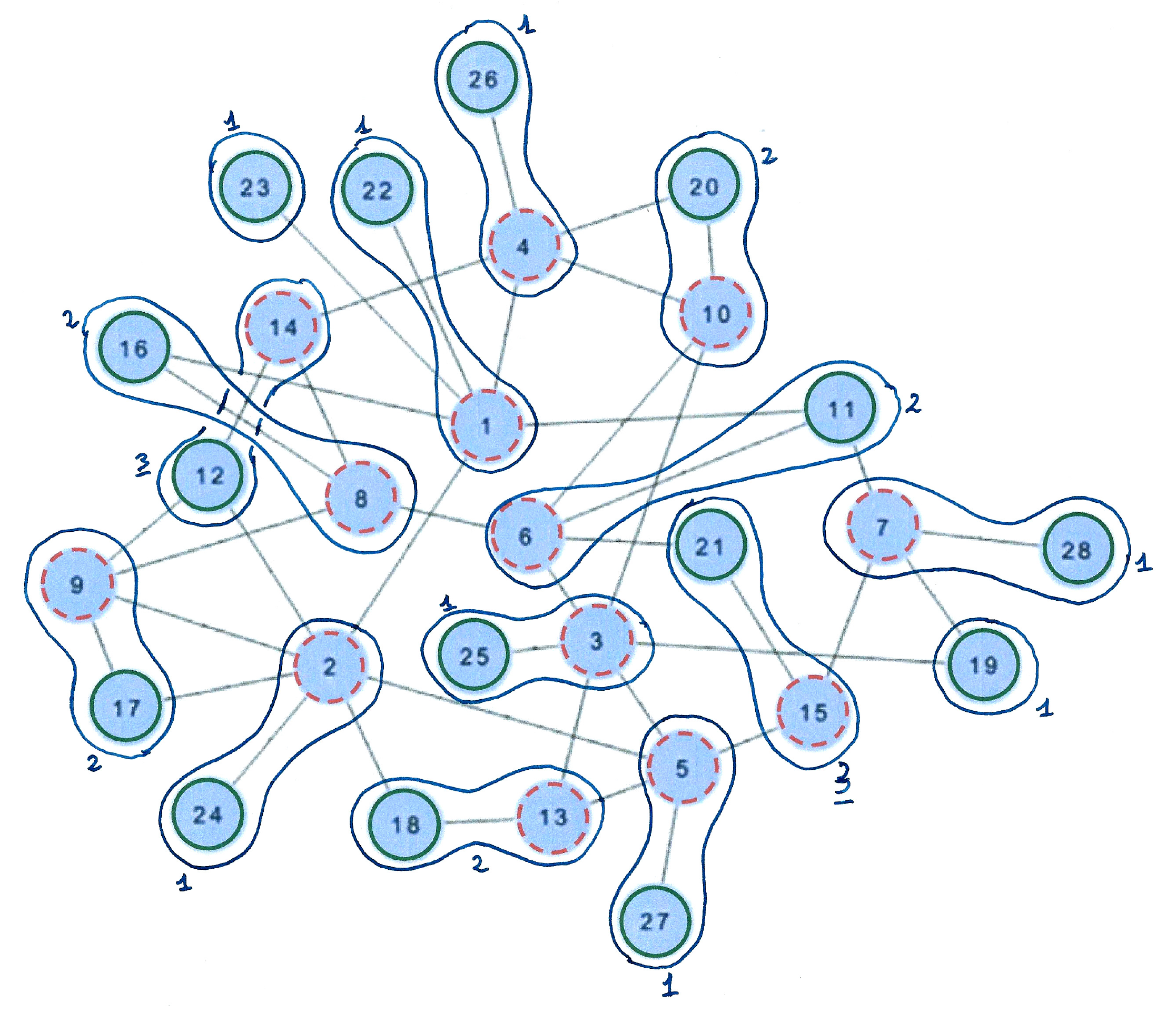} \\[-0.4cm]
  \caption{Graph of the random network with $N=28$ here investigated
and extracted from Ref.~\cite{Sev16}. Nodes are grouped by pairs (13) except 
nodes 19 and 23 which are left unpaired. Nodes encircled in green are sensors 
and those with red dashed contours are not. The rank $l$ of the layer for 
constructing the observer is also reported.}
  \label{topelec}
\end{figure}

We illustrate the pairing procedure for the network used in 
Ref.~\onlinecite{Sev16} and sketched in Fig.~\ref{topelec}: its size is $N=28$ 
and its average degree is $\langle k\rangle = 2.5$. There, paired nodes are
encircled in blue, node sensors have the green contour and those nodes
which must be fully reconstructed from another one have the red dashed contour. 
Note that some sensors are not paired (nodes 19 and 23). The algorithm 
producing this pairing is as follows:

\begin{enumerate}

\item Search for the nodes whose degree is equal to one and make a pair
with their only neighbor. If two nodes with degree $k=1$ have the same
neighbor (as nodes 22 and 23), then one of them is left unpaired (in
the example is node 23).

\item Search for the nodes with $k=2$ and pair them with an unpaired 
neighbor with the lowest degree.

\item Repeat step 2 increasing the degree and stop up to the stage at which
all nodes which can be paired are paired.

\end{enumerate}

As a result of applying these steps, we end up with 13 pairs and 2 unpaired 
nodes, that is, $N_m=15$ sensor nodes ($\frac{N}{2}+1$). In 
particular, this means that we have to measure $m = N$ variables (the $x_j$
and $y_j$ variables in the paired sensors and the $y_k$ variable in the 
unpaired ones). To check whether this choice provides a global observability, 
we have to compute the symbolic observability coefficient of the reconstructed 
vector whose components are
$\mb{X}_i = \left[ \begin{array}{ccc} y_i & \dot{y}_i & \ddot{y}_i 
\end{array} \right]^{\T}$ for the unpaired nodes $i=19,23$ and 
$\mb{X}_i = \left[ \begin{array}{cccccc} x_i & \dot{x}_i & y_i & \dot{y}_i & 
\ddot{y}_i & \stackrel{...}{y}_i \end{array} \right]^{\T}$ for the
paired sensors $i=11,12,16,17,18,20,21,22,24,25,26,27,$ and $28$. 
The symbolic observability coefficient is indeed equal to one and the
analytical determinant of the observability matrix is
Det~${\cal O}_X = \rho^{39}$, therefore validating our selection.
The expression of this determinant would mean that the observability is 
strongly sensitive to the coupling value. Nevertheless, since nodes are 
grouped by pairs, the dependency on the coupling value should not be 
practically worst than the one observed for a pair of nodes, that is, 
depending on $\rho^3$.

\begin{figure}
  \centering
  \includegraphics[width=0.3\textwidth]{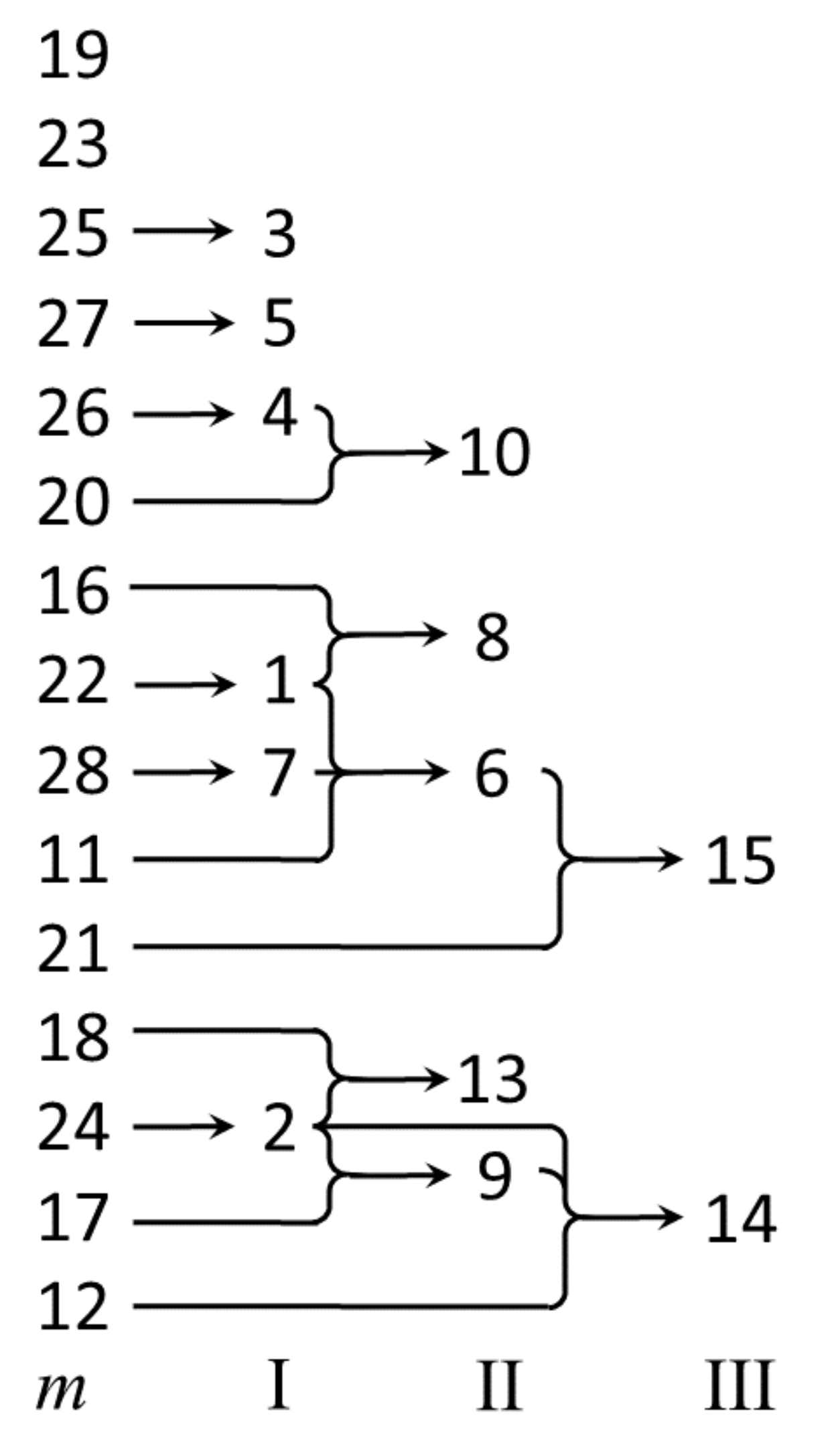} \\[-0.3cm]
  \caption{Hierarchical dependence of the observer. Nodal dynamics of the 
network shown in Fig.~\ref{topelec} is reconstructed layer by layer with the 
observer given by Eqs.~(\ref{obsdyadnet}) and (\ref{obssingle}). Layer 1 is 
composed exclusively by sensors. The reconstruction of the nodes belonging to 
layer 2 depends on those sensors and allow the reconstruction of those from 
layer 3, and so on.} 
  \label{classobserver}
\end{figure}

Before actually applying the network observer described by
Eqs.~(\ref{obsdyadnet}) for each pair $(i,j)$ and Eqs.~(\ref{obssingle})
for each unpaired sensor node, let us show how the reconstruction
works, especially regarding the coupling term in Eq.~(\ref{obsdyadnet}). For 
instance, to reconstruct node $i=8$ which is paired to the sensor $j=16$, we 
need to  subtract from the variable $y_{16}$ the signal $y_{1}$ from its 
neighbor (see Fig.\ \ref{topelec}). But, to know this coupling signal,
the dynamics of node $1$ needs to be reconstructed from node $22$,
which is a sensor, before completing the reconstruction process. This 
hierarchical relationship is depicted in Fig.\ \ref{classobserver}.  From left to right, we uncover four different layers 
composed of nodes whose reconstruction depend on the precedent layers. In the
first column of Fig.\ \ref{classobserver}, the sensor nodes (paired and 
unpaired) are the nodes which do not need any other nodes to be fully 
reconstructed: they belong to the layer 1. The reconstruction of the dynamics 
of nodes belonging to layer 2 are obtained from the nodal dynamics in layer 
1, and so on.

\begin{figure}
  \centering
  \begin{tabular}{c}
	  \includegraphics[width=0.34\textwidth]{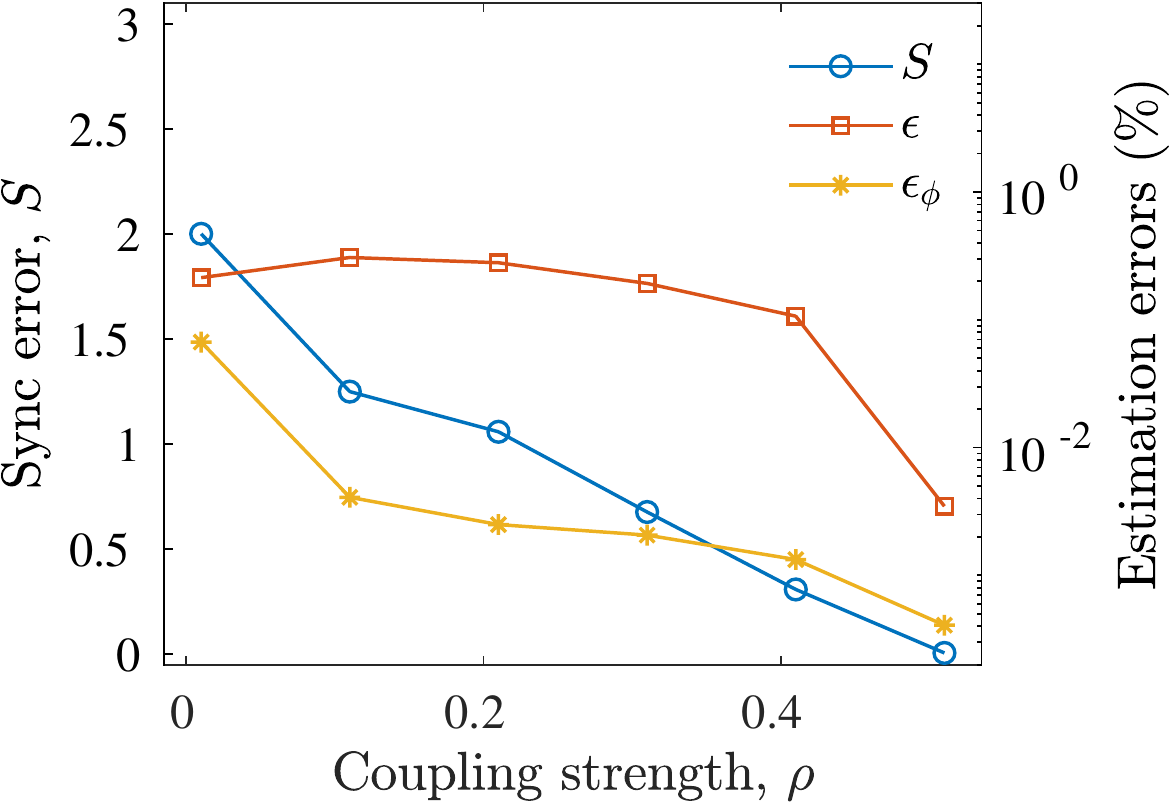} \\
	  (a) Identical oscillators: $a_i = 0.432$ \\[0.3cm]
 \includegraphics[width=0.34\textwidth]{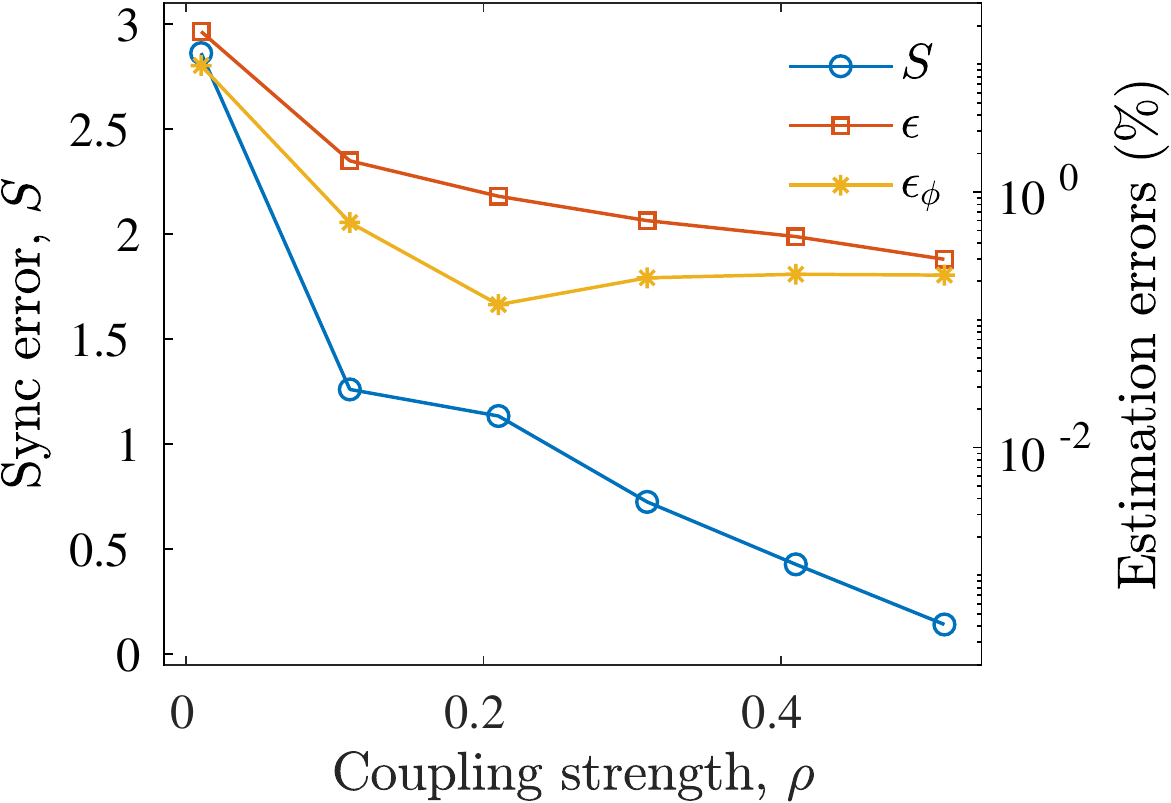} \\
	  (b) Non-identical oscillators: $a_i = 0.432 \pm 5\%$ \\[-0.2cm]
  \end{tabular}
  \caption{Small random network reconstruction. Synchronization error $S$, 
amplitude error $\epsilon$, and phase reconstruction error $\epsilon_\phi$ as a 
function of the coupling $\rho$ for the small random network with $N=28$ 
sketched in Fig.~\ref{topelec} for (a) identical  
and (b) non-identical oscillators.
In each case, the synchronization error $S$ is read in the left vertical axis 
while the reconstruction errors (in percentage) are read in the right vertical 
	axis. Each point is the average of 10 different initial conditions.}
  \label{ecircuit-recons}
\end{figure}

Once we solved the list of dependencies to reconstruct each node in our
network shown in Fig.~\ref{topelec}, we investigate the performance of the 
state observer by monitoring the error percentage in estimating the phase and 
amplitude of the nodal dynamics in the route to synchronization as  the 
coupling parameter is increased. As expected, when nodes are identical ($a_i
=0.432, \, \forall i$), all the errors are of the same order and keep low (less
than $0.3\%$) as for dyads or triads of identical R\"ossler systems [Fig. 
\ref{ecircuit-recons}(a)]; the phase and amplitude errors converge to zero 
when the oscillators synchronize to the same dynamics. Remarkably, when we 
inject a $5\%$ of parameter 
mismatch among the oscillators, the errors in the state estimation increase up 
to $15\%$ when the coupling is very low but they drop significantly down
to the levels observed for identical systems when the network leaves
the incoherent regime and synchronization is incipient
($\rho>0.1$).

\begin{figure*}
  \centering
  \includegraphics[width=0.3\textwidth]{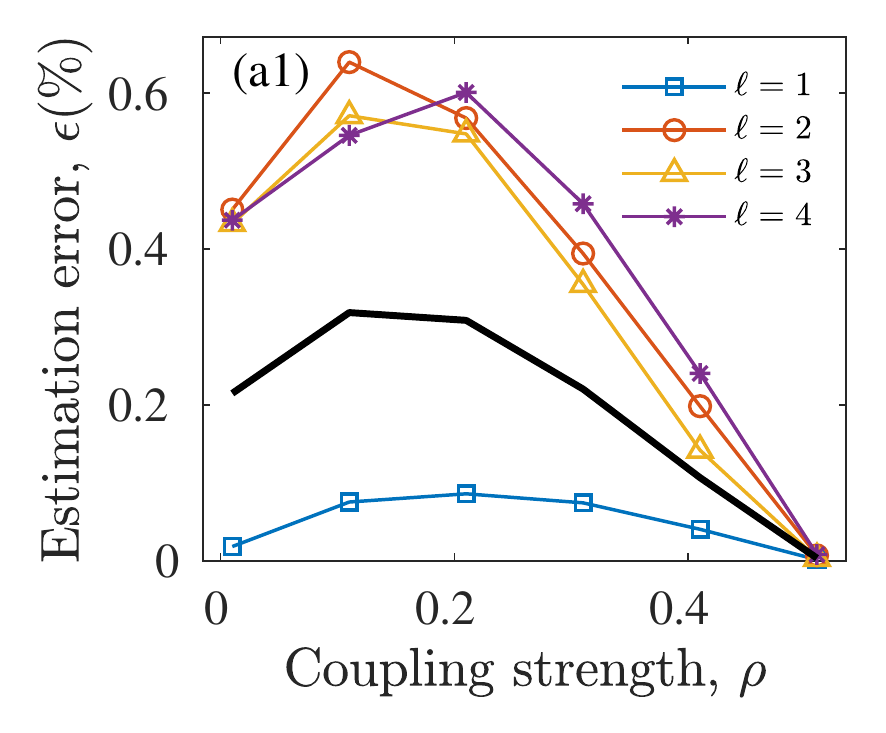}
 \includegraphics[width=0.3\textwidth]{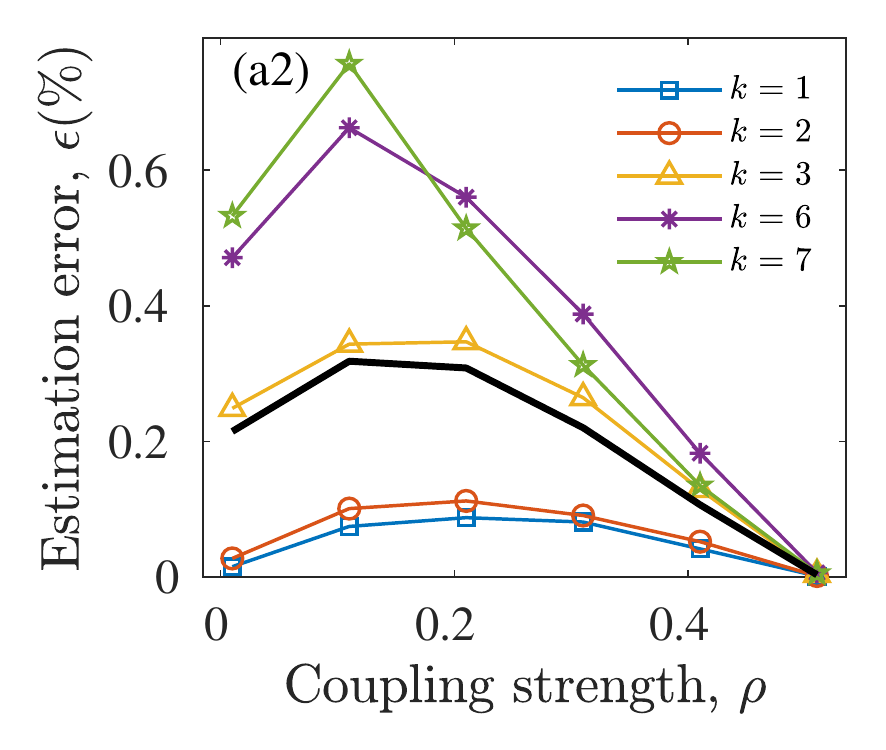}
  \includegraphics[width=0.3\textwidth]{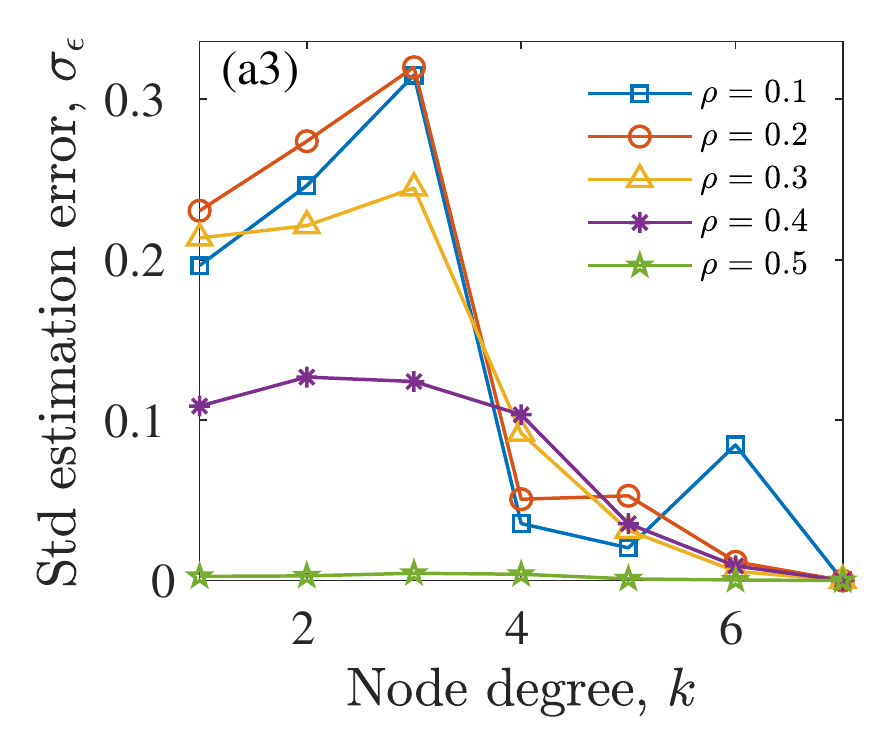}
	\\
	(a) Identical oscillators: $a_i = 0.432$ \\[0.2cm]
     \includegraphics[width=0.3\textwidth]{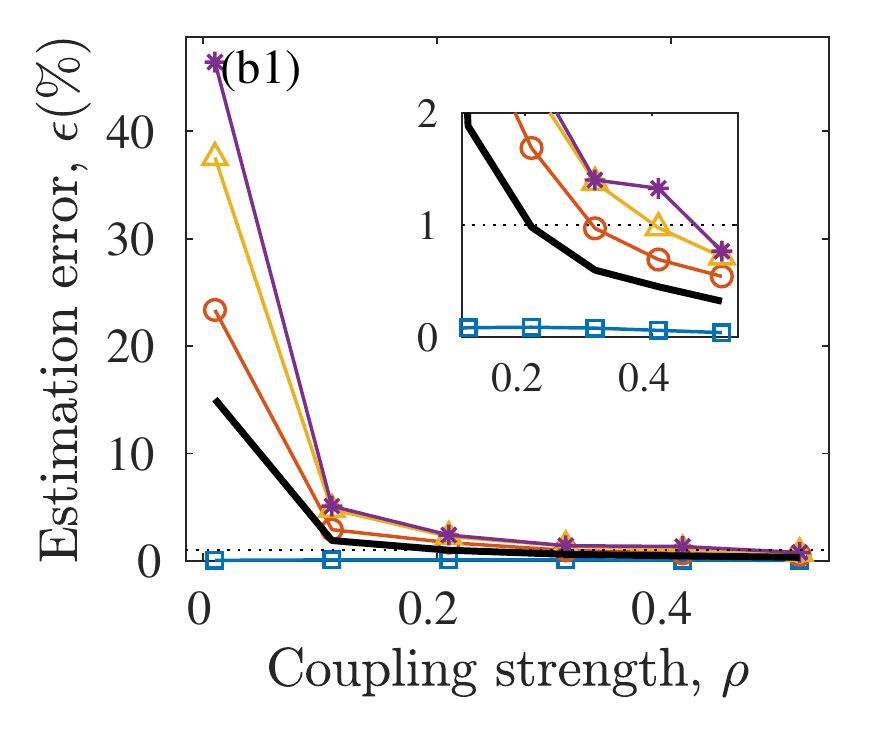}
        \includegraphics[width=0.3\textwidth]{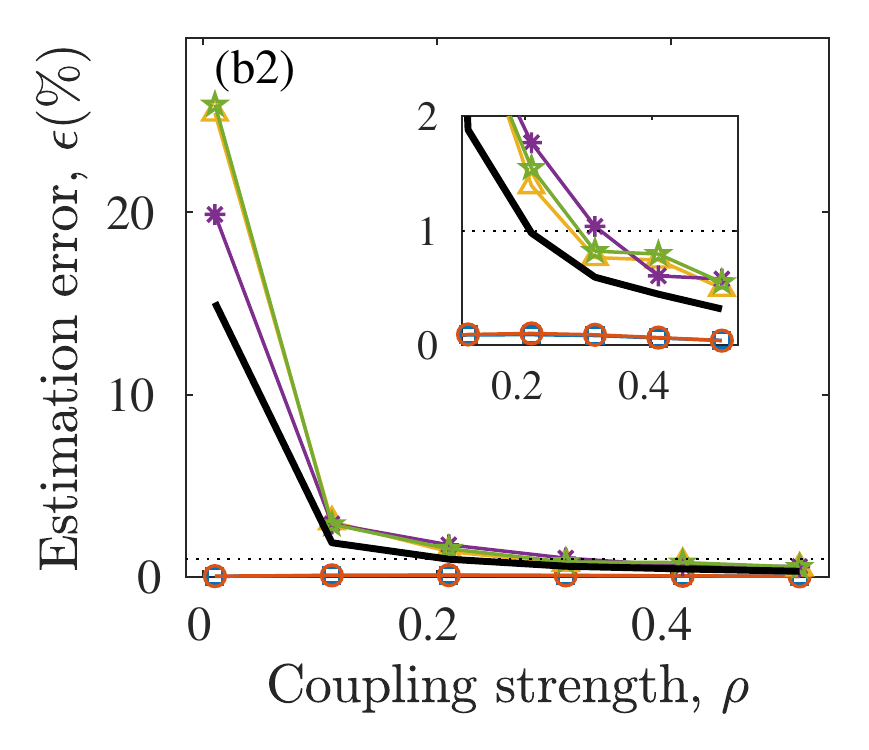}
 \includegraphics[width=0.3\textwidth]{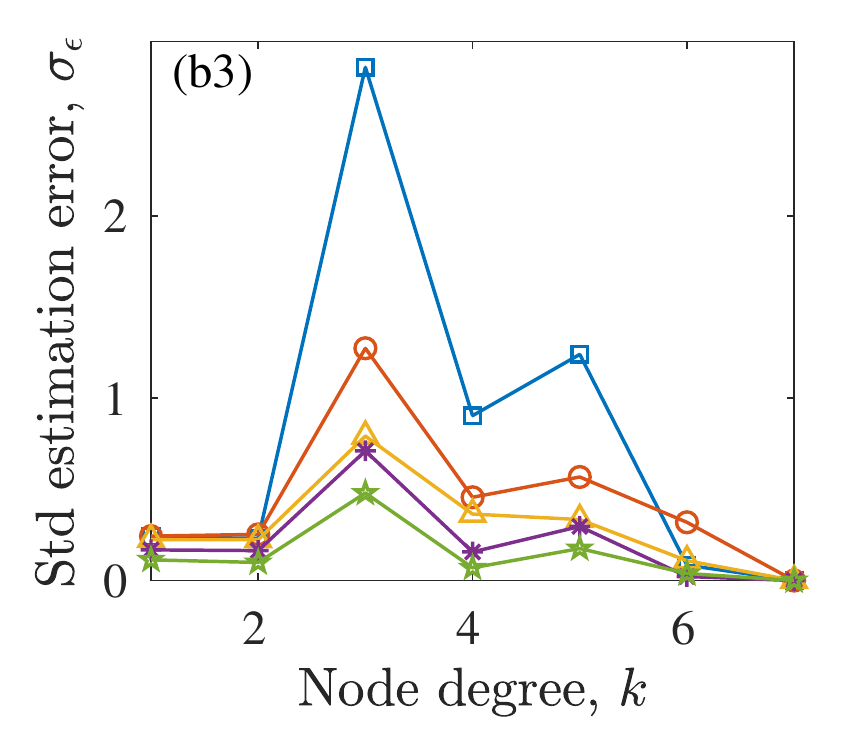} \\
	(b) Non-identical oscillators: $a_i = 0.432 \pm 5 \%$ \\[-0.2cm]
  \caption{Dependence of the estimation error $\epsilon$ on the reconstruction 
layer $l$ [(a1) and (b1)] and on the node degree $k$ [(a2) and (b2)] as a function of 
the coupling strength for the small network investigated in 
Fig.~\ref{ecircuit-recons}. The standard deviation of the estimation error 
within each node degree $k$ is shown for several coupling values in (a3) and (b3).  
Legends on top panels apply to the bottom ones. Insets are blow ups of the main 
plot. Black curves show average quantities. }
    \label{ecirc-recons-layer-classdegree} 
\end{figure*}

To explore in detail the role of the network topology in the network 
state reconstruction, we plot in Fig.~\ref{ecirc-recons-layer-classdegree} how 
the estimation error is distributed along the different layers and how it 
depends on the node degree $k$. Clearly, the best estimation is achieved in
the first layer as this is the sensing layer and the reconstruction of these 
nodes involves only one (the $\tilde{z}_j$ of a paired sensor) or two (the $\tilde{x}_k$ and $\tilde{z}_k$ of an unpaired sensor) variables, 
while in the rest of layers, the three dimensional dynamics of the nodes has to 
be fully reconstructed. Nevertheless, the performance of the state observer 
does not deteriorates as we go deeper in the layer structure, with all layers  
exhibiting similar errors and of the same order as the error of the first one 
[Fig.~\ref{ecirc-recons-layer-classdegree}(a1)]. This also holds for the 
network with non-identical oscillators 
[Fig.~\ref{ecirc-recons-layer-classdegree}(b1)] although here, for very low 
coupling regimes ($\rho < 0.1$), estimation errors are two orders of magnitude 
higher, with deeper layers having the largest errors. For $\rho > 0.1$, errors 
drop and keep below $2\%$ and within the same order for all layers (see the inset in Fig.\ref{ecirc-recons-layer-classdegree}(b1)). Figs.\ 
\ref{ecirc-recons-layer-classdegree}(a2) and 
\ref{ecirc-recons-layer-classdegree}(b2) show the same estimation errors as a 
function of the coupling strength but here each curve corresponds to the 
average error of all nodes having the same degree ranging between the minimum 
degree $k=1$, and the maximum $k=7$, 
being $k=3$ the average degree. Comparing 
\ref{ecirc-recons-layer-classdegree}(a2) and 
\ref{ecirc-recons-layer-classdegree}(b2), in both cases, the hierarchical 
procedure implemented in the reconstruction of the node dynamics is reflected 
here: nodes with degrees $k=1,2$ exhibit lower errors since those are most 
likely chosen as sensors, while those which are fully reconstructed belonging 
to deeper layers and having larger degrees, exhibit  estimation errors larger 
than the average (black curve) but still low. However, not all the nodes 
having the same degree show similar estimation errors as evidenced in Figs.\
\ref{ecirc-recons-layer-classdegree}(a3) and 
\ref{ecirc-recons-layer-classdegree}(b3) where the dispersion of the estimation 
error is plotted as a function of the node degree for different coupling 
values. Independently of the coupling strength, those nodes whose connectivity 
is close to the average degree of the network $\langle k \rangle \sim 3$, show 
the largest standard deviation, meaning that they are unevenly reconstructed 
with some of them being well predicted while others do not. 


\subsection{Role of the degree distribution}\label{numerical}

\begin{figure}
  \centering
  \includegraphics[width=0.3\textwidth]{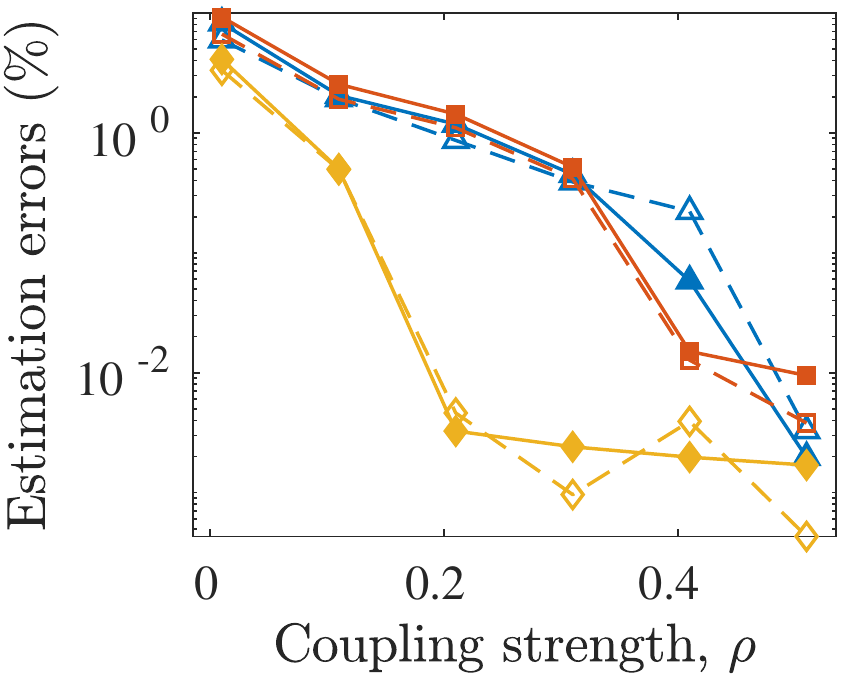} \\
	(a) Erd\"os-R\'enyi network \\[0.2cm]
  \includegraphics[width=0.3\textwidth]{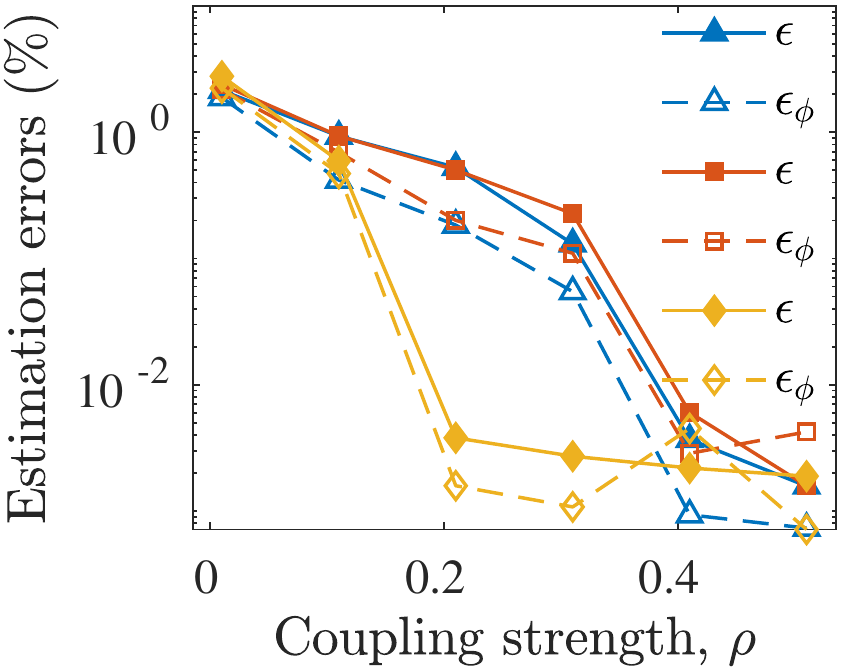}
	\\
	(b) Scale free network \\[-0.2cm]
     \caption{Estimation error of the amplitude (continuous lines) and of the 
phase (dashed lines) for ER (a) and SF (b) networks of size $N=128$ (triangles) 
and $N=256$ (squares) with $\langle k\rangle =4$ as a function of the coupling 
strength. The effect of increasing the mean degree for $N=128$ is also 
plotted for $\langle k\rangle =6$ (diamonds). Each point is the average of 10 
	different network realizations and initial conditions.}
  \label{ERSF-recons}
\end{figure}

To test the network observer in more complex connectivity architectures, we
use standard network models as the Erd\"os-R\'enyi (ER) model\cite{Erd59} for random graphs and the
 Barab\'asi-Albert model\cite{Bar99} to produce scale-free (SF) networks.   
We built network realizations as undirected graphs with $N = 128$ nodes and 
average connectivities $\langle k\rangle = 4$ and $\langle k\rangle =
6$. Figure \ref{ERSF-recons} summarizes the results for ER and SF networks 
showing that, in average, SF networks are easier to observe since the 
estimation errors in both phase and amplitude are lower: for intermediate 
coupling values, errors are around $1\%$ while for SF networks, they are about 
the half. Second, the network observer provides better estimations
when topologies are more densely connected, as shown in both panels of
Fig.~\ref{ERSF-recons} where curves for $\langle k\rangle =4$ are
above the ones for $\langle k\rangle =6$. We also checked the effect
of increasing the network size with no significant differences. 

\begin{figure}
  \centering
  \includegraphics[width=0.3\textwidth]{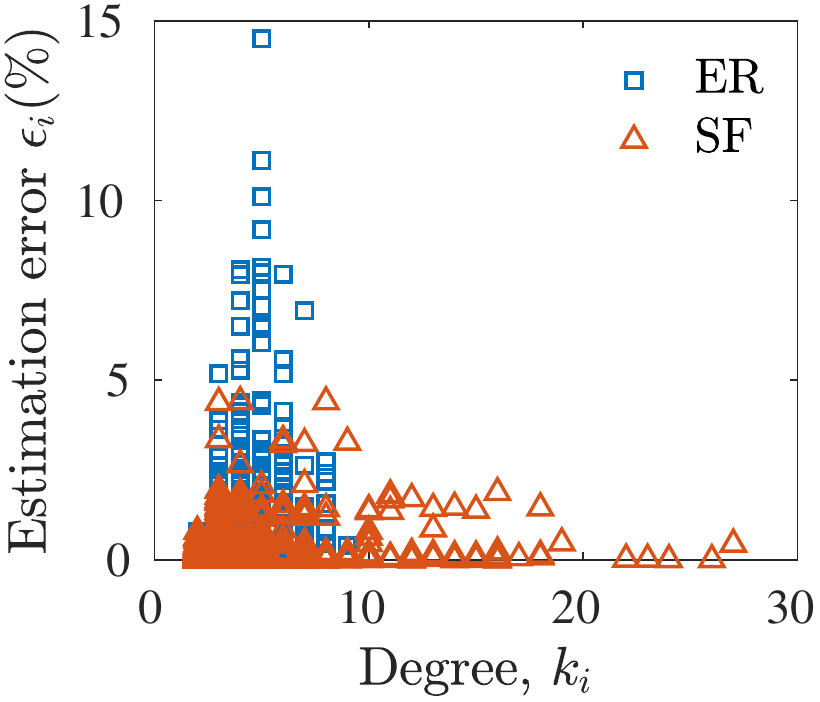}
	\\ (a) Estimation error \\[0.2cm]
  \includegraphics[width=0.3\textwidth]{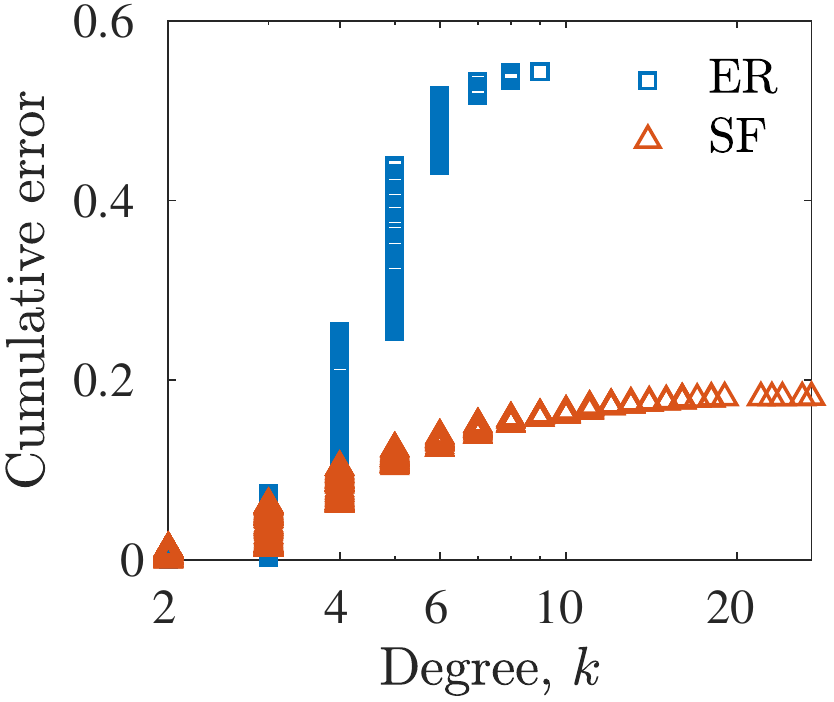}
	\\ (b) Cumulative estimation error \\[-0.2cm]
  \caption{Distribution of the estimation error of each node $\epsilon_i$ (a) 
and cumulative estimation error per node (b) as a function of the node degree 
for ER (squares) and SF (triangles) networks of size $N=128$ and $\langle
k\rangle=4$. Notice in panel (b) the log scale of the horizontal axis. 
Data correspond to the averaged values plotted in Fig.~\ref{ERSF-recons} at 
	$\rho = 0.2$. }
    \label{ERSF-recons-errdegs}
\end{figure}

The fact that the network state observer works better for SF than for ER
networks lies, indeed, in the distinct degree distributions featured by these 
topologies. Figure \ref{ERSF-recons-errdegs} shows, for a particular 
coupling value and $\langle k\rangle =4$, {how the estimation error in the 
reconstruction of the $N=128$ nodes, in each one of the 10 network realizations 
performed in Fig.~\ref{ERSF-recons}, distributes as a function of the node 
connectivity $k_i$}. Again, as in the example of the network with $N = 28$ 
nodes shown in Figs.\ \ref{ecirc-recons-layer-classdegree}(a3) and 
\ref{ecirc-recons-layer-classdegree}(b3), the nodes whose connectivity is 
around the mean connectivity $\langle k\rangle=4$ are those exhibiting a 
larger variety of estimation errors. This is clear in Fig.\ 
\ref{ERSF-recons-errdegs}(a) for both ER and SF networks. However, in the 
latter case, the small-degree nodes are the most numerous coexisting with a few 
highly connected hubs and, therefore, in comparison, the population of nodes 
with degree within the average is more reduced. An alternative way to inspect 
this is to plot how much error in the network state estimation is accumulated 
at nodes with degree smaller than a given $k$ 
[Fig.~\ref{ERSF-recons-errdegs}(b)]: it confirms the different contribution of 
the nodes depending on their connectivity. 

\begin{figure}
  \centering
  \includegraphics[width=0.234\textwidth]{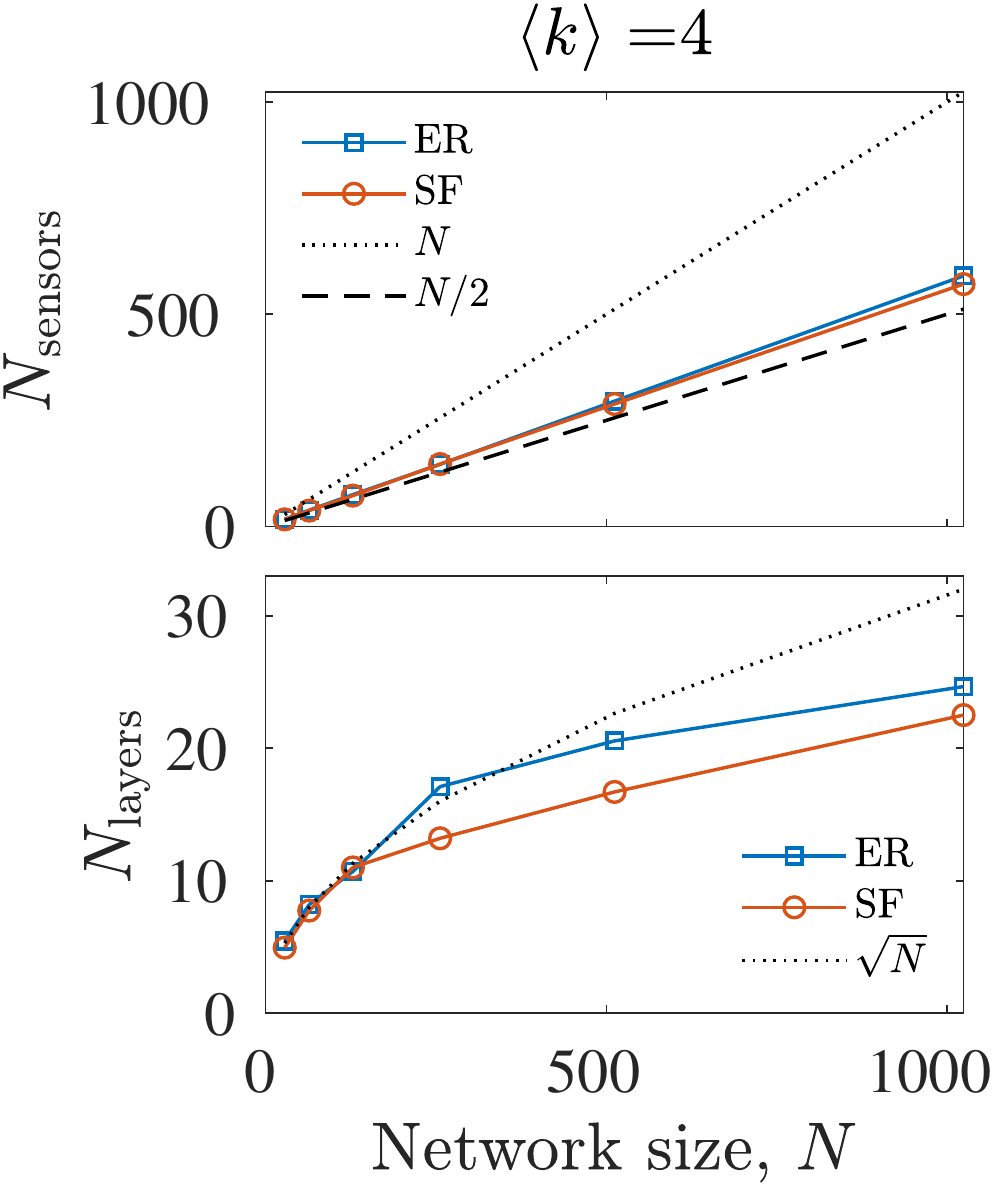}
  \includegraphics[width=0.234\textwidth]{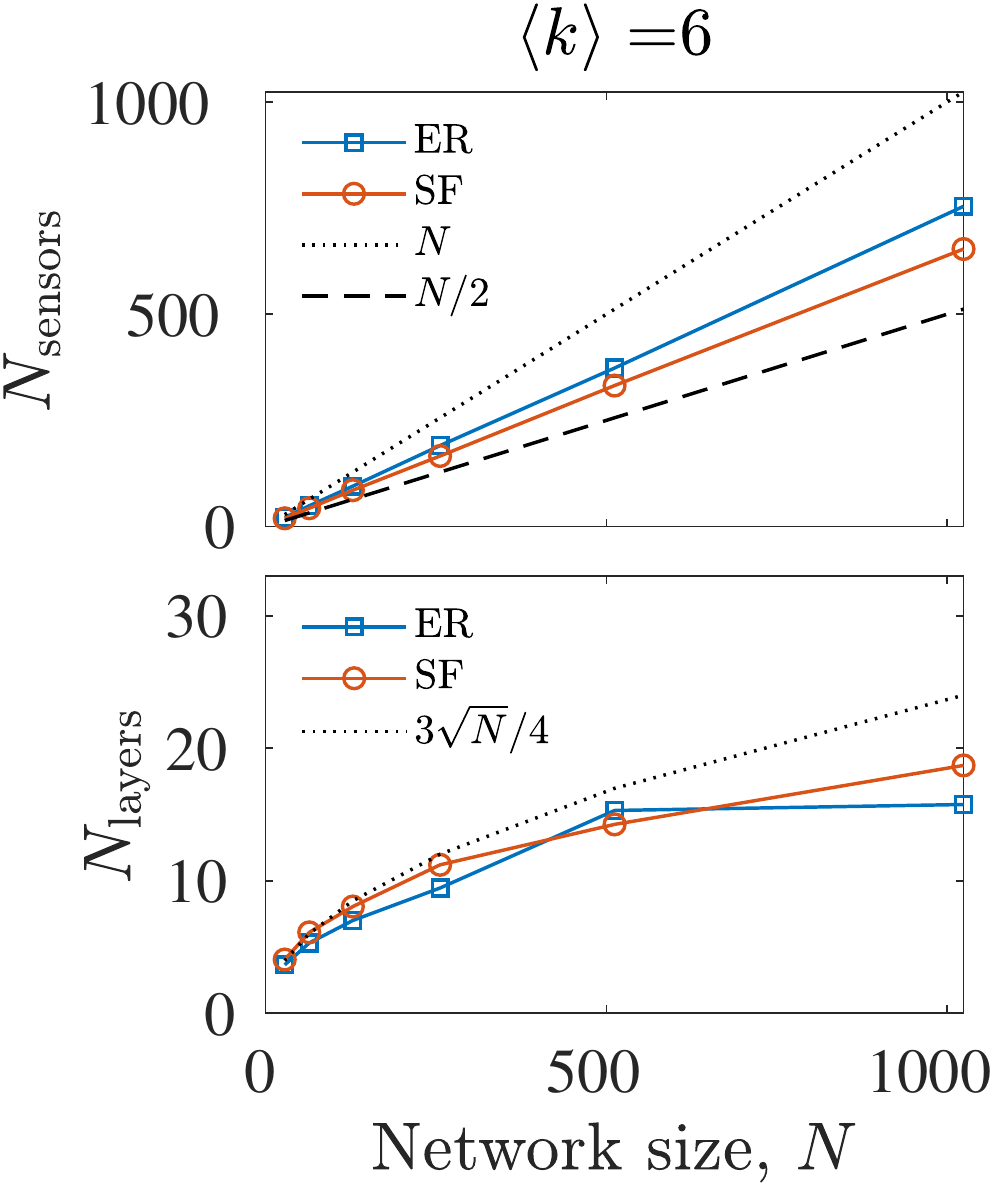}
  \caption{Number of sensors $N_{\rm sensors}$ and number of layers 
$N_{\rm layers}$ as a function of the network size ($N = \{ 28, 64, 128, 256, 
512, 1024 \}$) for (left column) $\langle k\rangle=4$ and (right column) $\langle k\rangle=6$. In this 
figure, each point is an average of 20 networks.}
  \label{NsNl_k}
\end{figure}

It is interesting to remark that while the number $N_{\rm sensors}$ of sensors is 
linearly correlated with the number $N$ of nodes, this is not the case for the 
number $N_{\rm layers}$ of layers which scales with $\sqrt{N}$ (Fig.\ \ref{NsNl_k}). 
Regarding $N_{\rm sensors}$, the  slope  is close to 0.5, slightly 
increasing as the mean degree $\langle k\rangle$ increases (compare top panels 
for $\langle k\rangle=4$ and $\langle k\rangle= 6$). As for the hierarchical 
reconstruction, the number $N_{\rm layers}$ of layers required is reduced as the 
average network connectivity is increased, saturating for large $N$. This would explain why the estimation errors are lower for networks with  larger average degree (see Fig.~\ref{ERSF-recons}).

\section{Conclusion}

We developed an algorithm to optimally place a set of sensors to render a network observable and designed
an observer returning its real time state. Our procedure is based
on an observability analysis adequate for nonlinear networks (here, the nodal
dynamics is nonlinear). Although the coupling function between nodes is linear, it is found
that the observer must be constructed in a layer by layer non-trivial way.
The key point  is to pair the nodes according to some basic topological rules
which can be automatized for treating arbitrarily large 
networks. We found that a network can be observed with errors which are of the 
same order of magnitude as those yield by dyads or triads of nodes. The errors become large in networks with heterogeneous node dynamics as long as the synchronization level remains low.
The performance of our network observer in the phase and amplitude reconstruction is very robust against different 
network architectures (degree distribution, link density and network size)
and parameter mismatch in the node dynamics, opening, therefore, the perspective to devise a robust control law for networks, subject which is currently under 
investigation.

\acknowledgements
This work was supported by the Ministerio de Econom\'ia, Industria y
Competitividad of Spain under project FIS2017-84151-P and Ministerio
de Ciencia e Innovaci\'on under project  PID2020-113737GB-I00.

\appendix

\bibliography{SysDyn}

\begin{thebibliography}{53}%
\makeatletter
\providecommand \@ifxundefined [1]{%
 \@ifx{#1\undefined}
}%
\providecommand \@ifnum [1]{%
 \ifnum #1\expandafter \@firstoftwo
 \else \expandafter \@secondoftwo
 \fi
}%
\providecommand \@ifx [1]{%
 \ifx #1\expandafter \@firstoftwo
 \else \expandafter \@secondoftwo
 \fi
}%
\providecommand \natexlab [1]{#1}%
\providecommand \enquote  [1]{``#1''}%
\providecommand \bibnamefont  [1]{#1}%
\providecommand \bibfnamefont [1]{#1}%
\providecommand \citenamefont [1]{#1}%
\providecommand \href@noop [0]{\@secondoftwo}%
\providecommand \href [0]{\begingroup \@sanitize@url \@href}%
\providecommand \@href[1]{\@@startlink{#1}\@@href}%
\providecommand \@@href[1]{\endgroup#1\@@endlink}%
\providecommand \@sanitize@url [0]{\catcode `\\12\catcode `\$12\catcode
  `\&12\catcode `\#12\catcode `\^12\catcode `\_12\catcode `\%12\relax}%
\providecommand \@@startlink[1]{}%
\providecommand \@@endlink[0]{}%
\providecommand \url  [0]{\begingroup\@sanitize@url \@url }%
\providecommand \@url [1]{\endgroup\@href {#1}{\urlprefix }}%
\providecommand \urlprefix  [0]{URL }%
\providecommand \Eprint [0]{\href }%
\providecommand \doibase [0]{http://dx.doi.org/}%
\providecommand \selectlanguage [0]{\@gobble}%
\providecommand \bibinfo  [0]{\@secondoftwo}%
\providecommand \bibfield  [0]{\@secondoftwo}%
\providecommand \translation [1]{[#1]}%
\providecommand \BibitemOpen [0]{}%
\providecommand \bibitemStop [0]{}%
\providecommand \bibitemNoStop [0]{.\EOS\space}%
\providecommand \EOS [0]{\spacefactor3000\relax}%
\providecommand \BibitemShut  [1]{\csname bibitem#1\endcsname}%
\let\auto@bib@innerbib\@empty
\bibitem [{\citenamefont {Levering}\ \emph {et~al.}(2013)\citenamefont
  {Levering}, \citenamefont {Kummer}, \citenamefont {Becker},\ and\
  \citenamefont {Sahle}}]{Lev13}%
  \BibitemOpen
  \bibfield  {author} {\bibinfo {author} {\bibfnamefont {J.}~\bibnamefont
  {Levering}}, \bibinfo {author} {\bibfnamefont {U.}~\bibnamefont {Kummer}},
  \bibinfo {author} {\bibfnamefont {K.}~\bibnamefont {Becker}}, \ and\ \bibinfo
  {author} {\bibfnamefont {S.}~\bibnamefont {Sahle}},\ }\bibfield  {title}
  {\enquote {\bibinfo {title} {Glycolytic oscillations in a model of a lactic
  acid bacterium metabolism},}\ }\href {\doibase 10.1016/j.bpc.2012.11.002}
  {\bibfield  {journal} {\bibinfo  {journal} {Biophysical Chemistry}\ }\textbf
  {\bibinfo {volume} {172}},\ \bibinfo {pages} {53--60} (\bibinfo {year}
  {2013})}\BibitemShut {NoStop}%
\bibitem [{\citenamefont {Srivastava}\ \emph {et~al.}(2020)\citenamefont
  {Srivastava}, \citenamefont {Nozari}, \citenamefont {Kim}, \citenamefont
  {Ju}, \citenamefont {Zhou}, \citenamefont {Becker}, \citenamefont
  {Pasqualetti}, \citenamefont {Pappas},\ and\ \citenamefont
  {Bassett}}]{Sri20}%
  \BibitemOpen
  \bibfield  {author} {\bibinfo {author} {\bibfnamefont {P.}~\bibnamefont
  {Srivastava}}, \bibinfo {author} {\bibfnamefont {E.}~\bibnamefont {Nozari}},
  \bibinfo {author} {\bibfnamefont {J.~Z.}\ \bibnamefont {Kim}}, \bibinfo
  {author} {\bibfnamefont {H.}~\bibnamefont {Ju}}, \bibinfo {author}
  {\bibfnamefont {D.}~\bibnamefont {Zhou}}, \bibinfo {author} {\bibfnamefont
  {C.}~\bibnamefont {Becker}}, \bibinfo {author} {\bibfnamefont
  {F.}~\bibnamefont {Pasqualetti}}, \bibinfo {author} {\bibfnamefont {G.~J.}\
  \bibnamefont {Pappas}}, \ and\ \bibinfo {author} {\bibfnamefont {D.~S.}\
  \bibnamefont {Bassett}},\ }\bibfield  {title} {\enquote {\bibinfo {title}
  {Models of communication and control for brain networks: distinctions,
  convergence, and future outlook},}\ }\href {\doibase 10.1162/netn_a_00158}
  {\bibfield  {journal} {\bibinfo  {journal} {Network Neuroscience}\ }\textbf
  {\bibinfo {volume} {4}},\ \bibinfo {pages} {1122--1159} (\bibinfo {year}
  {2020})}\BibitemShut {NoStop}%
\bibitem [{\citenamefont {Meng}\ and\ \citenamefont {Grebogi}(2021)}]{Meng21}%
  \BibitemOpen
  \bibfield  {author} {\bibinfo {author} {\bibfnamefont {Y.}~\bibnamefont
  {Meng}}\ and\ \bibinfo {author} {\bibfnamefont {C.}~\bibnamefont {Grebogi}},\
  }\bibfield  {title} {\enquote {\bibinfo {title} {Control of tipping points in
  stochastic mutualistic complex networks},}\ }\href {\doibase
  10.1063/5.0036051} {\bibfield  {journal} {\bibinfo  {journal} {Chaos}\
  }\textbf {\bibinfo {volume} {31}},\ \bibinfo {pages} {023118} (\bibinfo
  {year} {2021})}\BibitemShut {NoStop}%
\bibitem [{\citenamefont {Kalman}(1959)}]{Kal59}%
  \BibitemOpen
  \bibfield  {author} {\bibinfo {author} {\bibfnamefont {R.}~\bibnamefont
  {Kalman}},\ }\bibfield  {title} {\enquote {\bibinfo {title} {On the general
  theory of control systems},}\ }\href {\doibase 10.1109/TAC.1959.1104873}
  {\bibfield  {journal} {\bibinfo  {journal} {IRE Transactions on Automatic
  Control}\ }\textbf {\bibinfo {volume} {4}},\ \bibinfo {pages} {110--110}
  (\bibinfo {year} {1959})}\BibitemShut {NoStop}%
\bibitem [{\citenamefont {Hermann}\ and\ \citenamefont {Krener}(1977)}]{Her77}%
  \BibitemOpen
  \bibfield  {author} {\bibinfo {author} {\bibfnamefont {R.}~\bibnamefont
  {Hermann}}\ and\ \bibinfo {author} {\bibfnamefont {A.}~\bibnamefont
  {Krener}},\ }\bibfield  {title} {\enquote {\bibinfo {title} {Nonlinear
  controllability and observability},}\ }\href {\doibase
  10.1109/TAC.1977.1101601} {\bibfield  {journal} {\bibinfo  {journal} {IEEE
  Transactions on Automatic Control}\ }\textbf {\bibinfo {volume} {22}},\
  \bibinfo {pages} {728--740} (\bibinfo {year} {1977})}\BibitemShut {NoStop}%
\bibitem [{\citenamefont {Friedland}(1975)}]{Fri75}%
  \BibitemOpen
  \bibfield  {author} {\bibinfo {author} {\bibfnamefont {B.}~\bibnamefont
  {Friedland}},\ }\bibfield  {title} {\enquote {\bibinfo {title}
  {Controllability index based on conditioning number},}\ }\href {\doibase
  10.1115/1.3426963} {\bibfield  {journal} {\bibinfo  {journal} {Journal of
  Dynamic Systems, Measurement, and Control}\ }\textbf {\bibinfo {volume}
  {97}},\ \bibinfo {pages} {444--445} (\bibinfo {year} {1975})}\BibitemShut
  {NoStop}%
\bibitem [{\citenamefont {Aguirre}\ and\ \citenamefont
  {Letellier}(2005)}]{Agu05}%
  \BibitemOpen
  \bibfield  {author} {\bibinfo {author} {\bibfnamefont {L.~A.}\ \bibnamefont
  {Aguirre}}\ and\ \bibinfo {author} {\bibfnamefont {C.}~\bibnamefont
  {Letellier}},\ }\bibfield  {title} {\enquote {\bibinfo {title} {Observability
  of multivariate differential embeddings},}\ }\href@noop {} {\bibfield
  {journal} {\bibinfo  {journal} {Journal of Physics A: Mathematical and
  General}\ }\textbf {\bibinfo {volume} {38}},\ \bibinfo {pages} {6311}
  (\bibinfo {year} {2005})}\BibitemShut {NoStop}%
\bibitem [{\citenamefont {Letellier}\ \emph {et~al.}(1998)\citenamefont
  {Letellier}, \citenamefont {Maquet}, \citenamefont {Sceller}, \citenamefont
  {Gouesbet},\ and\ \citenamefont {Aguirre}}]{Let98b}%
  \BibitemOpen
  \bibfield  {author} {\bibinfo {author} {\bibfnamefont {C.}~\bibnamefont
  {Letellier}}, \bibinfo {author} {\bibfnamefont {J.}~\bibnamefont {Maquet}},
  \bibinfo {author} {\bibfnamefont {L.~L.}\ \bibnamefont {Sceller}}, \bibinfo
  {author} {\bibfnamefont {G.}~\bibnamefont {Gouesbet}}, \ and\ \bibinfo
  {author} {\bibfnamefont {L.~A.}\ \bibnamefont {Aguirre}},\ }\bibfield
  {title} {\enquote {\bibinfo {title} {On the non-equivalence of observables in
  phase-space reconstructions from recorded time series},}\ }\href@noop {}
  {\bibfield  {journal} {\bibinfo  {journal} {Journal of Physics A}\ }\textbf
  {\bibinfo {volume} {31}},\ \bibinfo {pages} {7913--7927} (\bibinfo {year}
  {1998})}\BibitemShut {NoStop}%
\bibitem [{\citenamefont {Letellier}\ and\ \citenamefont
  {Aguirre}(2002)}]{Let02}%
  \BibitemOpen
  \bibfield  {author} {\bibinfo {author} {\bibfnamefont {C.}~\bibnamefont
  {Letellier}}\ and\ \bibinfo {author} {\bibfnamefont {L.~A.}\ \bibnamefont
  {Aguirre}},\ }\bibfield  {title} {\enquote {\bibinfo {title} {Investigating
  nonlinear dynamics from time series: The influence of symmetries and the
  choice of observables},}\ }\href {\doibase 10.1063/1.1487570} {\bibfield
  {journal} {\bibinfo  {journal} {Chaos}\ }\textbf {\bibinfo {volume} {12}},\
  \bibinfo {pages} {549--558} (\bibinfo {year} {2002})}\BibitemShut {NoStop}%
\bibitem [{\citenamefont {Letellier}\ and\ \citenamefont
  {Aguirre}(2009)}]{Let09}%
  \BibitemOpen
  \bibfield  {author} {\bibinfo {author} {\bibfnamefont {C.}~\bibnamefont
  {Letellier}}\ and\ \bibinfo {author} {\bibfnamefont {L.~A.}\ \bibnamefont
  {Aguirre}},\ }\bibfield  {title} {\enquote {\bibinfo {title} {Symbolic
  observability coefficients for univariate and multivariate analysis},}\
  }\href {\doibase 10.1103/PhysRevE.79.066210} {\bibfield  {journal} {\bibinfo
  {journal} {Physical Review E}\ }\textbf {\bibinfo {volume} {79}},\ \bibinfo
  {pages} {066210} (\bibinfo {year} {2009})}\BibitemShut {NoStop}%
\bibitem [{\citenamefont {Bianco-Martinez}, \citenamefont {Baptista},\ and\
  \citenamefont {Letellier}(2015)}]{Bia15}%
  \BibitemOpen
  \bibfield  {author} {\bibinfo {author} {\bibfnamefont {E.}~\bibnamefont
  {Bianco-Martinez}}, \bibinfo {author} {\bibfnamefont {M.~S.}\ \bibnamefont
  {Baptista}}, \ and\ \bibinfo {author} {\bibfnamefont {C.}~\bibnamefont
  {Letellier}},\ }\bibfield  {title} {\enquote {\bibinfo {title} {Symbolic
  computations of nonlinear observability},}\ }\href {\doibase
  10.1103/PhysRevE.91.062912} {\bibfield  {journal} {\bibinfo  {journal}
  {Physical Review E}\ }\textbf {\bibinfo {volume} {91}},\ \bibinfo {pages}
  {062912} (\bibinfo {year} {2015})}\BibitemShut {NoStop}%
\bibitem [{\citenamefont {Letellier}\ \emph {et~al.}(2018)\citenamefont
  {Letellier}, \citenamefont {Sendi{\~n}a-Nadal}, \citenamefont
  {Bianco-Martinez},\ and\ \citenamefont {Baptista}}]{Let18}%
  \BibitemOpen
  \bibfield  {author} {\bibinfo {author} {\bibfnamefont {C.}~\bibnamefont
  {Letellier}}, \bibinfo {author} {\bibfnamefont {I.}~\bibnamefont
  {Sendi{\~n}a-Nadal}}, \bibinfo {author} {\bibfnamefont {E.}~\bibnamefont
  {Bianco-Martinez}}, \ and\ \bibinfo {author} {\bibfnamefont {M.~S.}\
  \bibnamefont {Baptista}},\ }\bibfield  {title} {\enquote {\bibinfo {title} {A
  symbolic network-based nonlinear theory for dynamical systems
  observability},}\ }\href {\doibase 10.1038/s41598-018-21967-w} {\bibfield
  {journal} {\bibinfo  {journal} {Scientific Reports}\ }\textbf {\bibinfo
  {volume} {8}},\ \bibinfo {pages} {3785} (\bibinfo {year} {2018})}\BibitemShut
  {NoStop}%
\bibitem [{\citenamefont {Lin}(1974)}]{Lin74}%
  \BibitemOpen
  \bibfield  {author} {\bibinfo {author} {\bibfnamefont {C.-T.}\ \bibnamefont
  {Lin}},\ }\bibfield  {title} {\enquote {\bibinfo {title} {Structural
  controllability},}\ }\href {\doibase 10.1109/TAC.1974.1100557} {\bibfield
  {journal} {\bibinfo  {journal} {IEEE Transactions on Automatic Control}\
  }\textbf {\bibinfo {volume} {19}},\ \bibinfo {pages} {201--208} (\bibinfo
  {year} {1974})}\BibitemShut {NoStop}%
\bibitem [{\citenamefont {Liu}, \citenamefont {Slotine},\ and\ \citenamefont
  {Barab\'asi}(2011)}]{Liu11}%
  \BibitemOpen
  \bibfield  {author} {\bibinfo {author} {\bibfnamefont {Y.-Y.}\ \bibnamefont
  {Liu}}, \bibinfo {author} {\bibfnamefont {J.-J.}\ \bibnamefont {Slotine}}, \
  and\ \bibinfo {author} {\bibfnamefont {A.-L.}\ \bibnamefont {Barab\'asi}},\
  }\bibfield  {title} {\enquote {\bibinfo {title} {Controllability of complex
  networks},}\ }\href {\doibase 10.1038/nature10011} {\bibfield  {journal}
  {\bibinfo  {journal} {Nature}\ }\textbf {\bibinfo {volume} {473}},\ \bibinfo
  {pages} {167--173} (\bibinfo {year} {2011})}\BibitemShut {NoStop}%
\bibitem [{\citenamefont {Liu}, \citenamefont {Slotine},\ and\ \citenamefont
  {Barab{\'a}si}(2013)}]{Liu13}%
  \BibitemOpen
  \bibfield  {author} {\bibinfo {author} {\bibfnamefont {Y.-Y.}\ \bibnamefont
  {Liu}}, \bibinfo {author} {\bibfnamefont {J.-J.}\ \bibnamefont {Slotine}}, \
  and\ \bibinfo {author} {\bibfnamefont {A.-L.}\ \bibnamefont {Barab{\'a}si}},\
  }\bibfield  {title} {\enquote {\bibinfo {title} {Observability of complex
  systems},}\ }\href {\doibase 10.1073/pnas.1215508110} {\bibfield  {journal}
  {\bibinfo  {journal} {Proceedings of the National Academy of Sciences}\
  }\textbf {\bibinfo {volume} {110}},\ \bibinfo {pages} {2460--2465} (\bibinfo
  {year} {2013})}\BibitemShut {NoStop}%
\bibitem [{\citenamefont {Liu}\ and\ \citenamefont {Pan}(2015)}]{Liu15}%
  \BibitemOpen
  \bibfield  {author} {\bibinfo {author} {\bibfnamefont {X.}~\bibnamefont
  {Liu}}\ and\ \bibinfo {author} {\bibfnamefont {L.}~\bibnamefont {Pan}},\
  }\bibfield  {title} {\enquote {\bibinfo {title} {Identifying driver nodes in
  the human signaling network using structural controllability analysis},}\
  }\href {\doibase 10.1109/TCBB.2014.2360396} {\bibfield  {journal} {\bibinfo
  {journal} {IEEE/ACM Transactions on Computational Biology and
  Bioinformatics}\ }\textbf {\bibinfo {volume} {12}},\ \bibinfo {pages}
  {467--472} (\bibinfo {year} {2015})}\BibitemShut {NoStop}%
\bibitem [{\citenamefont {Aguirre}, \citenamefont {Portes},\ and\ \citenamefont
  {Letellier}(2018)}]{Agu18}%
  \BibitemOpen
  \bibfield  {author} {\bibinfo {author} {\bibfnamefont {L.~A.}\ \bibnamefont
  {Aguirre}}, \bibinfo {author} {\bibfnamefont {L.~L.}\ \bibnamefont {Portes}},
  \ and\ \bibinfo {author} {\bibfnamefont {C.}~\bibnamefont {Letellier}},\
  }\bibfield  {title} {\enquote {\bibinfo {title} {Structural, dynamical and
  symbolic observability: From dynamical systems to networks},}\ }\href@noop {}
  {\bibfield  {journal} {\bibinfo  {journal} {PLoS ONE}\ }\textbf {\bibinfo
  {volume} {13}},\ \bibinfo {pages} {e0206180} (\bibinfo {year}
  {2018})}\BibitemShut {NoStop}%
\bibitem [{\citenamefont {Montanari}\ and\ \citenamefont
  {Aguirre}(2020)}]{Mon20}%
  \BibitemOpen
  \bibfield  {author} {\bibinfo {author} {\bibfnamefont {A.~N.}\ \bibnamefont
  {Montanari}}\ and\ \bibinfo {author} {\bibfnamefont {L.~A.}\ \bibnamefont
  {Aguirre}},\ }\bibfield  {title} {\enquote {\bibinfo {title} {Observability
  of network systems: A critical review of recent results},}\ }\href {\doibase
  10.1007/s40313-020-00633-5} {\bibfield  {journal} {\bibinfo  {journal}
  {Journal of Control, Automation and Electrical Systems}\ } (\bibinfo {year}
  {2020}),\ 10.1007/s40313-020-00633-5}\BibitemShut {NoStop}%
\bibitem [{\citenamefont {Mason}(1953)}]{Mas53}%
  \BibitemOpen
  \bibfield  {author} {\bibinfo {author} {\bibfnamefont {S.~J.}\ \bibnamefont
  {Mason}},\ }\bibfield  {title} {\enquote {\bibinfo {title} {Feedback theory
  -- some properties of signal flow graphs},}\ }\href {\doibase
  10.1109/JRPROC.1953.274449} {\bibfield  {journal} {\bibinfo  {journal}
  {Proceedings of the IRE}\ }\textbf {\bibinfo {volume} {41}},\ \bibinfo
  {pages} {1144--1156} (\bibinfo {year} {1953})}\BibitemShut {NoStop}%
\bibitem [{\citenamefont {Motter}(2015)}]{Mot15}%
  \BibitemOpen
  \bibfield  {author} {\bibinfo {author} {\bibfnamefont {A.~E.}\ \bibnamefont
  {Motter}},\ }\bibfield  {title} {\enquote {\bibinfo {title}
  {Networkcontrology},}\ }\href {\doibase 10.1063/1.4931570} {\bibfield
  {journal} {\bibinfo  {journal} {Chaos}\ }\textbf {\bibinfo {volume} {25}},\
  \bibinfo {pages} {097621} (\bibinfo {year} {2015})}\BibitemShut {NoStop}%
\bibitem [{\citenamefont {Whalen}\ \emph {et~al.}(2015)\citenamefont {Whalen},
  \citenamefont {Brennan}, \citenamefont {Sauer},\ and\ \citenamefont
  {Schiff}}]{Wha15}%
  \BibitemOpen
  \bibfield  {author} {\bibinfo {author} {\bibfnamefont {A.~J.}\ \bibnamefont
  {Whalen}}, \bibinfo {author} {\bibfnamefont {S.~N.}\ \bibnamefont {Brennan}},
  \bibinfo {author} {\bibfnamefont {T.~D.}\ \bibnamefont {Sauer}}, \ and\
  \bibinfo {author} {\bibfnamefont {S.~J.}\ \bibnamefont {Schiff}},\ }\bibfield
   {title} {\enquote {\bibinfo {title} {Observability and controllability of
  nonlinear networks: The role of symmetry},}\ }\href {\doibase
  10.1103/PhysRevX.5.011005} {\bibfield  {journal} {\bibinfo  {journal}
  {Physical Review X}\ }\textbf {\bibinfo {volume} {5}},\ \bibinfo {pages}
  {011005} (\bibinfo {year} {2015})}\BibitemShut {NoStop}%
\bibitem [{\citenamefont {Haber}, \citenamefont {Molnar},\ and\ \citenamefont
  {Motter}(2018)}]{Hab18}%
  \BibitemOpen
  \bibfield  {author} {\bibinfo {author} {\bibfnamefont {A.}~\bibnamefont
  {Haber}}, \bibinfo {author} {\bibfnamefont {F.}~\bibnamefont {Molnar}}, \
  and\ \bibinfo {author} {\bibfnamefont {A.~E.}\ \bibnamefont {Motter}},\
  }\bibfield  {title} {\enquote {\bibinfo {title} {State observation and sensor
  selection for nonlinear networks},}\ }\href@noop {} {\bibfield  {journal}
  {\bibinfo  {journal} {IEEE Transactions on Control of Network Systems}\
  }\textbf {\bibinfo {volume} {5}},\ \bibinfo {pages} {694--708} (\bibinfo
  {year} {2018})}\BibitemShut {NoStop}%
\bibitem [{\citenamefont {Jiang}\ and\ \citenamefont {Lai}(2019)}]{Jia19}%
  \BibitemOpen
  \bibfield  {author} {\bibinfo {author} {\bibfnamefont {J.}~\bibnamefont
  {Jiang}}\ and\ \bibinfo {author} {\bibfnamefont {Y.-C.}\ \bibnamefont
  {Lai}},\ }\bibfield  {title} {\enquote {\bibinfo {title} {Irrelevance of
  linear controllability to nonlinear dynamical networks},}\ }\href@noop {}
  {\bibfield  {journal} {\bibinfo  {journal} {Nature Communications}\ }\textbf
  {\bibinfo {volume} {10}},\ \bibinfo {pages} {3961} (\bibinfo {year}
  {2019})}\BibitemShut {NoStop}%
\bibitem [{\citenamefont {{Haber}}\ \emph {et~al.}(2021)\citenamefont
  {{Haber}}, \citenamefont {{Nugroho}}, \citenamefont {{Torres}},\ and\
  \citenamefont {{Taha}}}]{Hab21}%
  \BibitemOpen
  \bibfield  {author} {\bibinfo {author} {\bibfnamefont {A.}~\bibnamefont
  {{Haber}}}, \bibinfo {author} {\bibfnamefont {S.~A.}\ \bibnamefont
  {{Nugroho}}}, \bibinfo {author} {\bibfnamefont {P.}~\bibnamefont {{Torres}}},
  \ and\ \bibinfo {author} {\bibfnamefont {A.~F.}\ \bibnamefont {{Taha}}},\
  }\bibfield  {title} {\enquote {\bibinfo {title} {Control node selection
  algorithm for nonlinear dynamic networks},}\ }\href {\doibase
  10.1109/LCSYS.2020.3019591} {\bibfield  {journal} {\bibinfo  {journal} {IEEE
  Control Systems Letters}\ }\textbf {\bibinfo {volume} {5}},\ \bibinfo {pages}
  {1195--1200} (\bibinfo {year} {2021})}\BibitemShut {NoStop}%
\bibitem [{\citenamefont {Luenberger}(1964)}]{Lue64}%
  \BibitemOpen
  \bibfield  {author} {\bibinfo {author} {\bibfnamefont {D.~G.}\ \bibnamefont
  {Luenberger}},\ }\bibfield  {title} {\enquote {\bibinfo {title} {Observing
  the state of a linear system},}\ }\href {\doibase 10.1109/TME.1964.4323124}
  {\bibfield  {journal} {\bibinfo  {journal} {IEEE Transactions on Military
  Electronics}\ }\textbf {\bibinfo {volume} {8}},\ \bibinfo {pages} {74--80}
  (\bibinfo {year} {1964})}\BibitemShut {NoStop}%
\bibitem [{\citenamefont {Besan\c{c}on}(2007)}]{Bes07}%
  \BibitemOpen
  \bibfield  {author} {\bibinfo {author} {\bibfnamefont {G.}~\bibnamefont
  {Besan\c{c}on}},\ }\bibfield  {title} {\enquote {\bibinfo {title} {Nonlinear
  observers and applications},}\ }in\ \href@noop {} {\emph {\bibinfo
  {booktitle} {Lecture Notes in Control and Information Sciences}}},\ Vol.\
  \bibinfo {volume} {363}\ (\bibinfo  {publisher} {Springer-Verlag Berlin
  Heidelberg},\ \bibinfo {year} {2007})\BibitemShut {NoStop}%
\bibitem [{\citenamefont {Boutat}\ and\ \citenamefont {Zheng}(2021)}]{Bou21}%
  \BibitemOpen
  \bibfield  {author} {\bibinfo {author} {\bibfnamefont {D.}~\bibnamefont
  {Boutat}}\ and\ \bibinfo {author} {\bibfnamefont {G.}~\bibnamefont {Zheng}},\
  }\bibfield  {title} {\enquote {\bibinfo {title} {Observer design for
  nonlinear dynamical systems},}\ }in\ \href {\doibase
  10.1007/978-3-030-73742-9} {\emph {\bibinfo {booktitle} {Lecture Notes in
  Control and Information Sciences}}},\ Vol.\ \bibinfo {volume} {487}\
  (\bibinfo  {publisher} {Springer, Cham},\ \bibinfo {year} {2021})\BibitemShut
  {NoStop}%
\bibitem [{\citenamefont {Yu}\ \emph {et~al.}(2007)\citenamefont {Yu},
  \citenamefont {Chen}, \citenamefont {Cao}, \citenamefont {L\"u},\ and\
  \citenamefont {Parlitz}}]{Wen07}%
  \BibitemOpen
  \bibfield  {author} {\bibinfo {author} {\bibfnamefont {W.}~\bibnamefont
  {Yu}}, \bibinfo {author} {\bibfnamefont {G.}~\bibnamefont {Chen}}, \bibinfo
  {author} {\bibfnamefont {J.}~\bibnamefont {Cao}}, \bibinfo {author}
  {\bibfnamefont {J.}~\bibnamefont {L\"u}}, \ and\ \bibinfo {author}
  {\bibfnamefont {U.}~\bibnamefont {Parlitz}},\ }\bibfield  {title} {\enquote
  {\bibinfo {title} {Parameter identification of dynamical systems from time
  series},}\ }\href {\doibase 10.1103/PhysRevE.75.067201} {\bibfield  {journal}
  {\bibinfo  {journal} {Physical Review E}\ }\textbf {\bibinfo {volume} {75}},\
  \bibinfo {pages} {067201} (\bibinfo {year} {2007})}\BibitemShut {NoStop}%
\bibitem [{\citenamefont {Anguelova}, \citenamefont {Karlsson},\ and\
  \citenamefont {Jirstrand}(2012)}]{Ang12}%
  \BibitemOpen
  \bibfield  {author} {\bibinfo {author} {\bibfnamefont {M.}~\bibnamefont
  {Anguelova}}, \bibinfo {author} {\bibfnamefont {J.}~\bibnamefont {Karlsson}},
  \ and\ \bibinfo {author} {\bibfnamefont {M.}~\bibnamefont {Jirstrand}},\
  }\bibfield  {title} {\enquote {\bibinfo {title} {Minimal output sets for
  identifiability},}\ }\href {\doibase 10.1016/j.mbs.2012.04.005} {\bibfield
  {journal} {\bibinfo  {journal} {Mathematical Biosciences}\ }\textbf {\bibinfo
  {volume} {239}},\ \bibinfo {pages} {139--153} (\bibinfo {year}
  {2012})}\BibitemShut {NoStop}%
\bibitem [{\citenamefont {Aguirre}\ and\ \citenamefont
  {Letellier}(2009)}]{Agu09}%
  \BibitemOpen
  \bibfield  {author} {\bibinfo {author} {\bibfnamefont {L.~A.}\ \bibnamefont
  {Aguirre}}\ and\ \bibinfo {author} {\bibfnamefont {C.}~\bibnamefont
  {Letellier}},\ }\bibfield  {title} {\enquote {\bibinfo {title} {Modeling
  nonlinear dynamics and chaos: {A} review},}\ }\href {\doibase
  doi:10.1155/2009/238960} {\bibfield  {journal} {\bibinfo  {journal}
  {Mathematical Problems in Engineering}\ }\textbf {\bibinfo {volume} {2009}},\
  \bibinfo {pages} {238960} (\bibinfo {year} {2009})}\BibitemShut {NoStop}%
\bibitem [{\citenamefont {Pathak}\ \emph {et~al.}(2017)\citenamefont {Pathak},
  \citenamefont {Lu}, \citenamefont {Hunt}, \citenamefont {Girvan},\ and\
  \citenamefont {Ott}}]{Pat17}%
  \BibitemOpen
  \bibfield  {author} {\bibinfo {author} {\bibfnamefont {J.}~\bibnamefont
  {Pathak}}, \bibinfo {author} {\bibfnamefont {Z.}~\bibnamefont {Lu}}, \bibinfo
  {author} {\bibfnamefont {B.~R.}\ \bibnamefont {Hunt}}, \bibinfo {author}
  {\bibfnamefont {M.}~\bibnamefont {Girvan}}, \ and\ \bibinfo {author}
  {\bibfnamefont {E.}~\bibnamefont {Ott}},\ }\bibfield  {title} {\enquote
  {\bibinfo {title} {Using machine learning to replicate chaotic attractors and
  calculate {L}yapunov exponents from data},}\ }\href {\doibase
  10.1063/1.5010300} {\bibfield  {journal} {\bibinfo  {journal} {Chaos}\
  }\textbf {\bibinfo {volume} {27}},\ \bibinfo {pages} {121102} (\bibinfo
  {year} {2017})}\BibitemShut {NoStop}%
\bibitem [{\citenamefont {R{\"o}ssler}(1976)}]{Ros76c}%
  \BibitemOpen
  \bibfield  {author} {\bibinfo {author} {\bibfnamefont {O.~E.}\ \bibnamefont
  {R{\"o}ssler}},\ }\bibfield  {title} {\enquote {\bibinfo {title} {An equation
  for continuous chaos},}\ }\href {\doibase 10.1016/0375-9601(76)90101-8}
  {\bibfield  {journal} {\bibinfo  {journal} {Physics Letters A}\ }\textbf
  {\bibinfo {volume} {57}},\ \bibinfo {pages} {397--398} (\bibinfo {year}
  {1976})}\BibitemShut {NoStop}%
\bibitem [{\citenamefont {Sendi{\~n}a-Nadal}\ and\ \citenamefont
  {Letellier}(2019)}]{Sen19}%
  \BibitemOpen
  \bibfield  {author} {\bibinfo {author} {\bibfnamefont {I.}~\bibnamefont
  {Sendi{\~n}a-Nadal}}\ and\ \bibinfo {author} {\bibfnamefont {C.}~\bibnamefont
  {Letellier}},\ }\bibfield  {title} {\enquote {\bibinfo {title} {Observability
  of dynamical networks from graphic and symbolic approaches},}\ }in\ \href
  {\doibase 10.1007/978-3-030-14459-3\_1} {\emph {\bibinfo {booktitle}
  {Complenet 2019}}},\ \bibinfo {series} {Springer Proceedings in Complexity},
  Vol.~\bibinfo {volume} {X},\ \bibinfo {editor} {edited by\ \bibinfo {editor}
  {\bibfnamefont {S.}~\bibnamefont {Cornelius}}, \bibinfo {editor}
  {\bibfnamefont {C.~G.}\ \bibnamefont {Martorell}}, \bibinfo {editor}
  {\bibfnamefont {J.}~\bibnamefont {G{\'o}mez-Garde{\~n}es}}, \ and\ \bibinfo
  {editor} {\bibfnamefont {B.}~\bibnamefont {Gon\c{c}alves}}}\ (\bibinfo
  {publisher} {Springer, Cham},\ \bibinfo {year} {2019})\ pp.\ \bibinfo {pages}
  {3--15}\BibitemShut {NoStop}%
\bibitem [{\citenamefont {Chan}\ and\ \citenamefont {Shachter}(1992)}]{Cha92}%
  \BibitemOpen
  \bibfield  {author} {\bibinfo {author} {\bibfnamefont {B.~Y.}\ \bibnamefont
  {Chan}}\ and\ \bibinfo {author} {\bibfnamefont {R.~D.}\ \bibnamefont
  {Shachter}},\ }\bibfield  {title} {\enquote {\bibinfo {title} {Structural
  controllability and observability in influence diagrams},}\ }in\ \href@noop
  {} {\emph {\bibinfo {booktitle} {Proceedings of the Eighth International
  Conference on Uncertainty in Artificial Intelligence}}},\ \bibinfo {series
  and number} {UAI'92}\ (\bibinfo  {publisher} {Morgan Kaufmann Publishers
  Inc.},\ \bibinfo {address} {San Francisco, CA, USA},\ \bibinfo {year}
  {1992})\ pp.\ \bibinfo {pages} {25--32}\BibitemShut {NoStop}%
\bibitem [{\citenamefont {Letellier}, \citenamefont {Sendi{\~n}a-Nadal},\ and\
  \citenamefont {Aguirre}(2018)}]{Let18b}%
  \BibitemOpen
  \bibfield  {author} {\bibinfo {author} {\bibfnamefont {C.}~\bibnamefont
  {Letellier}}, \bibinfo {author} {\bibfnamefont {I.}~\bibnamefont
  {Sendi{\~n}a-Nadal}}, \ and\ \bibinfo {author} {\bibfnamefont {L.~A.}\
  \bibnamefont {Aguirre}},\ }\bibfield  {title} {\enquote {\bibinfo {title} {A
  nonlinear graph-based theory for dynamical network observability},}\ }\href
  {\doibase 10.1103/PhysRevE.98.020303} {\bibfield  {journal} {\bibinfo
  {journal} {Physical Review E}\ }\textbf {\bibinfo {volume} {98}},\ \bibinfo
  {pages} {020303(R)} (\bibinfo {year} {2018})}\BibitemShut {NoStop}%
\bibitem [{\citenamefont {Bianchin}\ \emph {et~al.}(2017)\citenamefont
  {Bianchin}, \citenamefont {Frasca}, \citenamefont {Gasparri},\ and\
  \citenamefont {Pasqualetti}}]{Bia17}%
  \BibitemOpen
  \bibfield  {author} {\bibinfo {author} {\bibfnamefont {G.}~\bibnamefont
  {Bianchin}}, \bibinfo {author} {\bibfnamefont {P.}~\bibnamefont {Frasca}},
  \bibinfo {author} {\bibfnamefont {A.}~\bibnamefont {Gasparri}}, \ and\
  \bibinfo {author} {\bibfnamefont {F.}~\bibnamefont {Pasqualetti}},\
  }\bibfield  {title} {\enquote {\bibinfo {title} {The observability radius of
  networks},}\ }\href {\doibase 10.1109/TAC.2016.2608941} {\bibfield  {journal}
  {\bibinfo  {journal} {IEEE Transactions on Automatic Control}\ }\textbf
  {\bibinfo {volume} {62}},\ \bibinfo {pages} {3006--3013} (\bibinfo {year}
  {2017})}\BibitemShut {NoStop}%
\bibitem [{\citenamefont {Hasegawa}, \citenamefont {Takaguchi},\ and\
  \citenamefont {Masuda}(2013)}]{Has13}%
  \BibitemOpen
  \bibfield  {author} {\bibinfo {author} {\bibfnamefont {T.}~\bibnamefont
  {Hasegawa}}, \bibinfo {author} {\bibfnamefont {T.}~\bibnamefont {Takaguchi}},
  \ and\ \bibinfo {author} {\bibfnamefont {N.}~\bibnamefont {Masuda}},\
  }\bibfield  {title} {\enquote {\bibinfo {title} {Observability transitions in
  correlated networks},}\ }\href {\doibase 10.1103/PhysRevE.88.042809}
  {\bibfield  {journal} {\bibinfo  {journal} {Physical Review E}\ }\textbf
  {\bibinfo {volume} {88}},\ \bibinfo {pages} {042809} (\bibinfo {year}
  {2013})}\BibitemShut {NoStop}%
\bibitem [{\citenamefont {Van~Mieghem}\ and\ \citenamefont
  {Wang}(2009)}]{Mie09}%
  \BibitemOpen
  \bibfield  {author} {\bibinfo {author} {\bibfnamefont {P.}~\bibnamefont
  {Van~Mieghem}}\ and\ \bibinfo {author} {\bibfnamefont {H.}~\bibnamefont
  {Wang}},\ }\bibfield  {title} {\enquote {\bibinfo {title} {The observable
  part of a network},}\ }\href {\doibase 10.1109/TNET.2008.925089} {\bibfield
  {journal} {\bibinfo  {journal} {IEEE/ACM Transactions in Networks}\ }\textbf
  {\bibinfo {volume} {17}},\ \bibinfo {pages} {93--105} (\bibinfo {year}
  {2009})}\BibitemShut {NoStop}%
\bibitem [{\citenamefont {Takens}(1981)}]{Tak81}%
  \BibitemOpen
  \bibfield  {author} {\bibinfo {author} {\bibfnamefont {F.}~\bibnamefont
  {Takens}},\ }\bibfield  {title} {\enquote {\bibinfo {title} {Detecting
  strange attractors in turbulence},}\ }\href {\doibase 10.1007/BFb0091924}
  {\bibfield  {journal} {\bibinfo  {journal} {Lectures Notes in Mathematics}\
  }\textbf {\bibinfo {volume} {898}},\ \bibinfo {pages} {366--381} (\bibinfo
  {year} {1981})}\BibitemShut {NoStop}%
\bibitem [{\citenamefont {Letellier}, \citenamefont {Minati},\ and\
  \citenamefont {Barbot}(2022)}]{Let22a}%
  \BibitemOpen
  \bibfield  {author} {\bibinfo {author} {\bibfnamefont {C.}~\bibnamefont
  {Letellier}}, \bibinfo {author} {\bibfnamefont {L.}~\bibnamefont {Minati}}, \
  and\ \bibinfo {author} {\bibfnamefont {J.-P.}\ \bibnamefont {Barbot}},\
  }\bibfield  {title} {\enquote {\bibinfo {title} {Optimal placement of sensor
  and actuator for controlling the piecewise linear {Ch}ua circuit via a
  discrete-time observer},}\ }\href@noop {} {\bibfield  {journal} {\bibinfo
  {journal} {submitted}\ } (\bibinfo {year} {2022})}\BibitemShut {NoStop}%
\bibitem [{\citenamefont {Krener}\ and\ \citenamefont
  {Respondek}(1985)}]{Kre85}%
  \BibitemOpen
  \bibfield  {author} {\bibinfo {author} {\bibfnamefont {A.~J.}\ \bibnamefont
  {Krener}}\ and\ \bibinfo {author} {\bibfnamefont {W.}~\bibnamefont
  {Respondek}},\ }\bibfield  {title} {\enquote {\bibinfo {title} {Nonlinear
  observers with linearizable error dynamics},}\ }\href {\doibase
  10.1137/0323016} {\bibfield  {journal} {\bibinfo  {journal} {SIAM Journal on
  Control and Optimization}\ }\textbf {\bibinfo {volume} {23}},\ \bibinfo
  {pages} {197--216} (\bibinfo {year} {1985})}\BibitemShut {NoStop}%
\bibitem [{\citenamefont {Kailath}(1980)}]{Kai80}%
  \BibitemOpen
  \bibfield  {author} {\bibinfo {author} {\bibfnamefont {T.}~\bibnamefont
  {Kailath}},\ }\href@noop {} {\emph {\bibinfo {title} {Linear Systems}}},\
  Information and System Sciences Series\ (\bibinfo  {publisher}
  {Prentice-Hall},\ \bibinfo {year} {1980})\BibitemShut {NoStop}%
\bibitem [{\citenamefont {Letellier}, \citenamefont {Aguirre},\ and\
  \citenamefont {Maquet}(2005)}]{Let05a}%
  \BibitemOpen
  \bibfield  {author} {\bibinfo {author} {\bibfnamefont {C.}~\bibnamefont
  {Letellier}}, \bibinfo {author} {\bibfnamefont {L.~A.}\ \bibnamefont
  {Aguirre}}, \ and\ \bibinfo {author} {\bibfnamefont {J.}~\bibnamefont
  {Maquet}},\ }\bibfield  {title} {\enquote {\bibinfo {title} {Relation between
  observability and differential embeddings for nonlinear dynamics},}\ }\href
  {\doibase 10.1103/PhysRevE.71.066213} {\bibfield  {journal} {\bibinfo
  {journal} {Physical Review E}\ }\textbf {\bibinfo {volume} {71}},\ \bibinfo
  {pages} {066213} (\bibinfo {year} {2005})}\BibitemShut {NoStop}%
\bibitem [{\citenamefont {Sendi{\~n}a-Nadal}, \citenamefont {Boccaletti},\ and\
  \citenamefont {Letellier}(2016)}]{Sen16}%
  \BibitemOpen
  \bibfield  {author} {\bibinfo {author} {\bibfnamefont {I.}~\bibnamefont
  {Sendi{\~n}a-Nadal}}, \bibinfo {author} {\bibfnamefont {S.}~\bibnamefont
  {Boccaletti}}, \ and\ \bibinfo {author} {\bibfnamefont {C.}~\bibnamefont
  {Letellier}},\ }\bibfield  {title} {\enquote {\bibinfo {title} {Observability
  coefficients for predicting the class of synchronizability from the algebraic
  structure of the local oscillators},}\ }\href {\doibase
  10.1103/PhysRevE.94.042205} {\bibfield  {journal} {\bibinfo  {journal}
  {Physical Review E}\ }\textbf {\bibinfo {volume} {94}},\ \bibinfo {pages}
  {042205} (\bibinfo {year} {2016})}\BibitemShut {NoStop}%
\bibitem [{\citenamefont {Frunzete}, \citenamefont {Barbot},\ and\
  \citenamefont {Letellier}(2012)}]{Fru12}%
  \BibitemOpen
  \bibfield  {author} {\bibinfo {author} {\bibfnamefont {M.}~\bibnamefont
  {Frunzete}}, \bibinfo {author} {\bibfnamefont {J.-P.}\ \bibnamefont
  {Barbot}}, \ and\ \bibinfo {author} {\bibfnamefont {C.}~\bibnamefont
  {Letellier}},\ }\bibfield  {title} {\enquote {\bibinfo {title} {Influence of
  the singular manifold of nonobservable states in reconstructing chaotic
  attractors},}\ }\href {\doibase 10.1103/PhysRevE.86.026205} {\bibfield
  {journal} {\bibinfo  {journal} {Physical Review E}\ }\textbf {\bibinfo
  {volume} {86}},\ \bibinfo {pages} {026205} (\bibinfo {year}
  {2012})}\BibitemShut {NoStop}%
\bibitem [{\citenamefont {Osipov}\ \emph {et~al.}(2003)\citenamefont {Osipov},
  \citenamefont {Hu}, \citenamefont {Zhou}, \citenamefont {Ivanchenko},\ and\
  \citenamefont {Kurths}}]{Osi03}%
  \BibitemOpen
  \bibfield  {author} {\bibinfo {author} {\bibfnamefont {G.~V.}\ \bibnamefont
  {Osipov}}, \bibinfo {author} {\bibfnamefont {B.}~\bibnamefont {Hu}}, \bibinfo
  {author} {\bibfnamefont {C.}~\bibnamefont {Zhou}}, \bibinfo {author}
  {\bibfnamefont {M.~V.}\ \bibnamefont {Ivanchenko}}, \ and\ \bibinfo {author}
  {\bibfnamefont {J.}~\bibnamefont {Kurths}},\ }\bibfield  {title} {\enquote
  {\bibinfo {title} {Three types of transitions to phase synchronization in
  coupled chaotic oscillators},}\ }\href@noop {} {\bibfield  {journal}
  {\bibinfo  {journal} {Physical Review Letters}\ }\textbf {\bibinfo {volume}
  {91}},\ \bibinfo {pages} {024101} (\bibinfo {year} {2003})}\BibitemShut
  {NoStop}%
\bibitem [{\citenamefont {Nyquist}(1924)}]{Nyq24}%
  \BibitemOpen
  \bibfield  {author} {\bibinfo {author} {\bibfnamefont {H.}~\bibnamefont
  {Nyquist}},\ }\bibfield  {title} {\enquote {\bibinfo {title} {Certain factors
  affecting telegraph speed},}\ }\href {\doibase
  10.1002/j.1538-7305.1924.tb01361.x} {\bibfield  {journal} {\bibinfo
  {journal} {Bell System Technical Journal}\ }\textbf {\bibinfo {volume} {3}},\
  \bibinfo {pages} {324--346} (\bibinfo {year} {1924})}\BibitemShut {NoStop}%
\bibitem [{\citenamefont {Letellier}, \citenamefont {Aguirre},\ and\
  \citenamefont {Freitas}(2009)}]{Fre09b}%
  \BibitemOpen
  \bibfield  {author} {\bibinfo {author} {\bibfnamefont {C.}~\bibnamefont
  {Letellier}}, \bibinfo {author} {\bibfnamefont {L.~A.}\ \bibnamefont
  {Aguirre}}, \ and\ \bibinfo {author} {\bibfnamefont {U.~S.}\ \bibnamefont
  {Freitas}},\ }\bibfield  {title} {\enquote {\bibinfo {title} {Frequently
  asked questions about global modeling},}\ }\href {\doibase 10.1063/1.3125705}
  {\bibfield  {journal} {\bibinfo  {journal} {Chaos}\ }\textbf {\bibinfo
  {volume} {19}},\ \bibinfo {pages} {023103} (\bibinfo {year}
  {2009})}\BibitemShut {NoStop}%
\bibitem [{\citenamefont {Letellier}, \citenamefont {Dutertre},\ and\
  \citenamefont {Maheu}(1995)}]{Let95a}%
  \BibitemOpen
  \bibfield  {author} {\bibinfo {author} {\bibfnamefont {C.}~\bibnamefont
  {Letellier}}, \bibinfo {author} {\bibfnamefont {P.}~\bibnamefont {Dutertre}},
  \ and\ \bibinfo {author} {\bibfnamefont {B.}~\bibnamefont {Maheu}},\
  }\bibfield  {title} {\enquote {\bibinfo {title} {Unstable periodic orbits and
  templates of the {R}\"ossler system: {T}oward a systematic topological
  characterization},}\ }\href {\doibase 10.1063/1.166076} {\bibfield  {journal}
  {\bibinfo  {journal} {Chaos}\ }\textbf {\bibinfo {volume} {5}},\ \bibinfo
  {pages} {271--282} (\bibinfo {year} {1995})}\BibitemShut {NoStop}%
\bibitem [{\citenamefont {Mangiarotti}, \citenamefont {Sendi{\~n}a-Nadal},\
  and\ \citenamefont {Letellier}(2018)}]{Man18b}%
  \BibitemOpen
  \bibfield  {author} {\bibinfo {author} {\bibfnamefont {S.}~\bibnamefont
  {Mangiarotti}}, \bibinfo {author} {\bibfnamefont {I.}~\bibnamefont
  {Sendi{\~n}a-Nadal}}, \ and\ \bibinfo {author} {\bibfnamefont
  {C.}~\bibnamefont {Letellier}},\ }\bibfield  {title} {\enquote {\bibinfo
  {title} {Using global modeling to unveil hidden couplings in small network
  motifs},}\ }\href {\doibase 10.1063/1.5037335} {\bibfield  {journal}
  {\bibinfo  {journal} {Chaos}\ }\textbf {\bibinfo {volume} {28}},\ \bibinfo
  {pages} {123110} (\bibinfo {year} {2018})}\BibitemShut {NoStop}%
\bibitem [{\citenamefont {Sevilla-Escoboza}\ and\ \citenamefont
  {Buld{\'u}}(2016)}]{Sev16}%
  \BibitemOpen
  \bibfield  {author} {\bibinfo {author} {\bibfnamefont {R.}~\bibnamefont
  {Sevilla-Escoboza}}\ and\ \bibinfo {author} {\bibfnamefont {J.~M.}\
  \bibnamefont {Buld{\'u}}},\ }\bibfield  {title} {\enquote {\bibinfo {title}
  {Synchronization of networks of chaotic oscillators: Structural and dynamical
  datasets},}\ }\href {\doibase https://doi.org/10.1016/j.dib.2016.03.097}
  {\bibfield  {journal} {\bibinfo  {journal} {Data in Brief}\ }\textbf
  {\bibinfo {volume} {7}},\ \bibinfo {pages} {1185--1189} (\bibinfo {year}
  {2016})}\BibitemShut {NoStop}%
\bibitem [{\citenamefont {Erd{\H{o}}s}\ and\ \citenamefont
  {R{\'e}nyi}(1959)}]{Erd59}%
  \BibitemOpen
  \bibfield  {author} {\bibinfo {author} {\bibfnamefont {P.}~\bibnamefont
  {Erd{\H{o}}s}}\ and\ \bibinfo {author} {\bibfnamefont {A.}~\bibnamefont
  {R{\'e}nyi}},\ }\bibfield  {title} {\enquote {\bibinfo {title} {On random
  graphs},}\ }\href@noop {} {\bibfield  {journal} {\bibinfo  {journal}
  {Publicationes Mathematicae Debrecen}\ }\textbf {\bibinfo {volume} {6}},\
  \bibinfo {pages} {290--297} (\bibinfo {year} {1959})}\BibitemShut {NoStop}%
\bibitem [{\citenamefont {Barab{\'a}si}\ and\ \citenamefont
  {Albert}(1999)}]{Bar99}%
  \BibitemOpen
  \bibfield  {author} {\bibinfo {author} {\bibfnamefont {A.-L.}\ \bibnamefont
  {Barab{\'a}si}}\ and\ \bibinfo {author} {\bibfnamefont {R.}~\bibnamefont
  {Albert}},\ }\bibfield  {title} {\enquote {\bibinfo {title} {Emergence of
  scaling in random networks},}\ }\href@noop {} {\bibfield  {journal} {\bibinfo
   {journal} {Science}\ }\textbf {\bibinfo {volume} {286}},\ \bibinfo {pages}
  {509--512} (\bibinfo {year} {1999})}\BibitemShut {NoStop}%
\end{thebibliography}%
\end{document}